\documentclass[10pt,preprint]{aastex}

\slugcomment{To appear in the Astrophysical Journal}
\shortauthors{Fitzpatrick \& Massa}
\shorttitle{Extinction Survey}

\begin{document}

\newcommand{\atlas}{{ATLAS9}}
\newcommand{\tlusty}{{TLUSTY}}
\newcommand{\synspec}{{SYNSPEC}}
\newcommand{\iras}{{\it IRAS}}
\newcommand{\ans}{{\it ANS}}
\newcommand{\hst}{{\it HST}}
\newcommand{\iue}{{\it IUE}}
\newcommand{\oao}{{\it OAO-2}}
\newcommand{\td}{{\it TD-1}}
\newcommand{\tmass}{{2MASS}}                                                                                                                
\newcommand{\mast}{{MAST}}
\newcommand{\irsa}{{IRSA}}
\newcommand{\hip}{{\it Hipparcos}}
\newcommand{\simbad}{{\it SIMBAD}}

\newcommand{\teff}{\mbox{$T_{\rm eff}$}}
\newcommand{\logg}{{$\log g$}}
\newcommand{\vturb}{$v_{turb}$}
\newcommand{\abund}{[m/H]}
\newcommand{\vsini}{$v \sin i$}

\newcommand{\kms}{km\,s$^{-1}$}
\newcommand{\msun}{${\rm M}_\sun$}
\newcommand{\ebv}{\mbox{$E(B\!-\!V)$}}
\newcommand{\invmic}{\mbox{$\mu{\rm m}^{-1}$}}

\title{An Analysis of the Shapes of Interstellar Extinction Curves. V.
The IR-Through-UV Curve Morphology}

\author{E.L.~Fitzpatrick\altaffilmark{1}, D.~Massa\altaffilmark{2}}
\altaffiltext{1}{Department of Astronomy \& Astrophysics, Villanova
University, 800 Lancaster Avenue, Villanova, PA 19085, USA; 
fitz@astronomy.villanova.edu}
\altaffiltext{2}{SGT, Inc., NASA/GSFC, Mailstop 665.0, Greenbelt,
MD 20771; massa@derckmassa.net}

\begin{abstract}
We study the IR-through-UV wavelength dependence of 328 Galactic
interstellar extinction curves affecting normal, near-main sequence B
and late O stars.  We derive the curves using a new technique which
employs stellar atmosphere models in lieu of unreddened ``standard''
stars.  Under ideal conditions, this technique is capable of virtually
eliminating spectral mismatch errors in the curves.  In general, it
lends itself to a quantitative assessment of the errors and enables a
rigorous testing of the significance of relationships between various
curve parameters, regardless of whether their uncertainties are
correlated.  Analysis of the curves gives the following results:
\begin{enumerate}
\item In accord with our previous findings, the central position of the
2175 \AA\/ extinction bump is mildly variable, its width is highly
variable, and the two variations are unrelated.  \item Strong
correlations are found among some extinction properties within the UV
region, and within the IR region.
\item With the exception of a few curves with extreme (i.e., large) values of $R(V)$,
{\it the UV and IR portions of Galactic extinction curves are not
correlated with each other.}
\item The large sightline-to-sightline variation seen in our sample
implies that any average Galactic extinction curve will always reflect the
biases of its parent sample.  
\item The use of an average curve to deredden a spectral energy
distribution (SED) will result in significant errors, and a realistic
error budget for the dereddened SED must include the observed variance of
Galactic curves.  
\end{enumerate}
While the observed large sightline-to-sightline variations, and the
lack of correlation among the various features of the curves, make it
difficult to meaningfully characterize average extinction properties,
they demonstrate that extinction curves respond sensitively to local
conditions. Thus, each curve contains potentially unique information
about the grains along its sightline.
\end{abstract}



\section{INTRODUCTION\label{secINTRO}}
In the previous paper in this series (Fitzpatrick \& Massa 2005a,
hereafter Paper IV), we introduced a technique,
``extinction-without-standards,'' to determine the shapes of
UV-through-IR interstellar extinction curves by modeling the observed
spectral energy distributions (SEDs) of reddened early-type stars. The
method involves a $\chi^2$-minimization procedure to determine
simultaneously the basic properties of a reddened star (namely, \teff,
\logg, \abund, and \vturb) {\it and} the shape of its extinction curve,
utilizing grids of stellar atmosphere models to represent intrinsic
SEDs and an analytical form of the extinction curve, whose shape is
determined by a set of adjustable parameters.

In general, the benefits of extinction-without-standards are increased
accuracy and precision (in most applications) over results generated
using the standard Pair Method of extinction curve determination and,
very importantly, a reliable estimate of the uncertainties in the
resultant extinction curves.  Specifically, the advantages of the new
method include: (1) the elimination of the requirement for observations
of unreddened spectral standard stars, (2) the near-elimination of
``spectral mismatch'' as a source of extinction curve error, (3) the
ability to produce accurate curves for much more lightly-reddened
sightlines than heretofore possible, (4) the ability to derive accurate
ultraviolet curves for later spectral types (i.e., to late-B classes)
than previously possible, and (5) the ability to provide quantified
estimates of the degree of correlation between various morphological
features of the curves.

The chief limitation of the method is that the intrinsic SEDs of the
reddened stars must be well-represented by model atmosphere
calculations.  This currently eliminates from consideration such
objects as high luminosity stars, Be stars, Wolf-Rayet stars, and any
spectrally peculiar star.  This restriction also affects the Pair
Method, since there are only a small number of unreddened members of
these classes which can serve as ``standard stars,'' and it is
difficult to be certain that the intrinsic SEDs of the standard stars
really represent those of the reddened stars.  In addition, the extinction-without-standards technique requires well-calibrated (and absolutely-calibrated) SED observations.

In this paper, we apply our extinction-without-standards technique to a
sample of 328 Galactic stars for which multi-wavelength SEDs are
available.  For these stars, we derive normalized UV-through-IR
extinction curves, sets of parameters which describe the shapes of the
curves, and sets of parameters which characterize the stars
themselves.  The scope of our discussion focuses on two objectives: (1)
the presentation of the broad-ranging results and a description of the
methodology employed, and (2) a thorough examination of general
extinction curve morphology.  Among the issues addressed is the
correlation between the IR and UV properties of extinction (see, e.g.,
Cardelli, Clayton, \& Mathis 1989).  This paper is not intended as a
review of Galactic extinction and we explicitly restrict our attention
only to issues touched upon by our new analysis.  The results, however,
do constitute a broad view of Galactic extinction and lend themselves
to numerous other investigations, including the general properties of
extinction curves, regional trends in extinction properties, the
correlation between extinction and other interstellar properties,
determination of intrinsic SEDs for objects in clusters containing
survey stars, the study of small scale spatial variations in dust grain
populations from stars in cluster extinction curves, etc. We plan to
pursue some of these in future papers.  Preliminary results from an
early version of this study were reported by Fitzpatrick (2004;
hereafter F04), which also provided a review of then-recent progress in
interstellar extinction studies.
  
The sample of stars chosen for this survey and the data used are
described in \S \ref{secDATA}.  This is followed in \S \ref{secMETHOD}
by a  brief description of the extinction-without-curves technique,
including a discussion of the error analysis --- which is critical to
the analyses in the latter parts of the paper.  The results of the
survey, essentially an atlas of Galactic extinction curves, is
presented in \S \ref{secATLAS}, with the full sets of tables and the
full set of figures from this section available in the electronic
edition of the Journal.  In \S \ref{secSTARS} we briefly discuss the
stellar properties derived from our analysis and then in \S
\ref{secEXTINCTION} present a detailed description of Galactic
extinction curve morphology, from the IR to the UV spectral regions.
Finally, in \S \ref{secDISCUSS} we provide a brief summary of the chief
conclusions of this study.


\section{THE SURVEY STARS AND THEIR DATA\label{secDATA}}

In principle, the extinction-without-standards technique can be applied
to, or expanded to include, any type of SED data.  In practice,
however, we have developed the technique with specific datasets in
mind, namely, low-resolution UV spectrophotometry from the {\it
International Ultraviolet Explorer} (\iue) satellite and ground-based
optical and near-IR photometry.  In this particular study, we utilize
\iue\/ spectrophotometry, {\it UBV} photometry, and {\it JHK}
photometry from the {Two-Micron All Sky Survey} (\tmass).  Although
some of the survey stars have additional data available, e.g.,
Str\"{o}mgren $ uvby\beta$ photometry, we elected --- for the sake of
uniformity --- to include only the {\it UBV} data in the analysis.  As
a result, the stellar parameters we derive in this paper will be less
accurate than those determined in Paper IV.  Nevertheless, the errors
in these parameters are still well-determined.

The most restrictive of the datasets is that of the \iue, and so we
began our search for survey stars with the \iue\/ database.  Using the
search engine provided by the {Multimission Archive at STScI} (\mast),
we examined all low-resolution spectra for stars with \iue\/ Object
Class numbers of 12 (main sequence O), 20 (B0-B2 V-IV), 21 (B3-B5
V-IV), 22 (B6-B9.5 V-IV), 23 (B0-B2 III-I), 24 (B3-B5 III-I), and 25
(B6-B9.5 III-I).  Because our goal is to obtain a uniform dataset of
reddened stars whose SEDs can be modeled accurately, we eliminated the
following types of objects from the available field of candidates: (1)
stars without good-quality spectra from both the short-wavelength (SWP)
and long-wavelength (LWR or LWP) \iue\/ spectral regions, (2)
clearly-unreddened stars, (3) known Be stars, (4) luminosity class I
stars, (5) O stars more luminous than class V, (6) O stars earlier than
spectral type O5, and (7) stars with peculiar-looking UV spectra (as
based on our own assessment).

The list of potential candidates from the \iue\/ database was then
examined for the availability of {\it UBV} data, using the General
Catalog of Photometric Data (GCPD) maintained at the University of
Geneva (see Mermilliod, Mermilliod, \& Hauck 1997)\footnotemark
\footnotetext{The GCPD catalog was accessed via the website at
http://obswww.unige.ch/gcpd/gcpd.html.}.  In general, stars without
{\it V} and \bv\/ measurements were eliminated from consideration.
However, broadband Geneva photometry was available for five stars
without {\it UBV} data (BD+56 526, HD62542, HD108927, HD110336, and
HD143054), and so we included these stars and utilized the Geneva $U-B$
and $V-B$ indices.

The final trimming of the survey sample was not performed until the
SEDs had been modeled. At this point we imposed the requirement that
all survey stars must have values of $\ebv \geq 0.20$ mag.  This limit
is somewhat arbitrary and, as shown by Paper IV, useful extinction
curves can be derived via the extinction-without-standards technique
for \ebv\/ values considerably lower than 0.20 mag.  However, the
uncertainties do rise at low \ebv\/ and for this survey we wanted a
sample of stars for which the uncertainties in the final parameters
were uniformly small.  We plan in the future to examine the
extinction properties along lightly reddened sightlines.

Near-IR {\it JHK} photometry (and their associated uncertainties) were
retrieved for the survey stars from the \tmass\/ database at the
NASA/IPAC Infrared Science Archive (\irsa)\footnotemark
\footnotetext{The \tmass\/ data were accessed via the \irsa\/ website
at http://irsa.ipac.caltech.edu/applications/Gator.}.  Only those
\tmass\/ magnitudes for which the uncertainties are less than $\pm0.1$
mag are used here.  For seven stars whose \tmass\/ measurements were
either non-existent or of low quality (HD23180, HD23512, HD37022,
HD144470, HD147165, HD147933, and HD149757), Johnson {\it JHK}
magnitudes were available and were retrieved via the GCPD catalog.
Note that the availability of {\it JHK} data was not a requirement for
inclusion in this survey although, as will be quantified below, most of
our sample have such data.

The last data collection activity was to retrieve ancillary information
for all stars, such as coordinates, alternate names, and spectral
types, using the \simbad\/ database.

The only data processing required for this program involved the \iue\/
spectrophotometry.  The \iue\/ data contained in the \mast\/ archive
were processed using the NEWSIPS software (Nichols \& Linsky 1996).  As
discussed in detail by Massa and Fitzpatrick (2000; hereafter MF00),
these data contain significant thermal and temporal dependencies and
suffer from an incorrect absolute calibration.  We corrected the data
for their systematic errors and placed them onto the \hst/FOS flux
scale of Bohlin (1996) using the corrections and algorithms described
and provided by MF00.  This step is absolutely essential for our
program since our ``comparison stars'' for deriving extinction curves
are stellar atmosphere models and systematic errors in the absolute
calibration of the data do not cancel out as they would in the case of
the Pair Method.  (Note that the thermally- and temporally-dependent
errors in the NEWSIPS data would not generally cancel out in the Pair
Method --- see MF00.)  When multiple spectra were available in one of
\iue{\it 's} wavelength ranges (SWP or LWR and LWP), they were combined
using the NEWSIPS error arrays as weights.  Small aperture data were
scaled to the large aperture data and both trailed and point source
data were included.  Short and long wavelength data were joined at
1978~\AA\ to form a complete spectrum covering the wavelength range
$1150 \leq \lambda \leq 3000$~\AA.  Data longward of 3000~\AA\ were
ignored because they are typically of low quality and subject to
residual systematic effects.

After all the limits and restrictions were imposed, we arrived at our
final sample of 328 stars for which UV spectrophotometry covering
1150-3000~\AA\/ and {\it UBV} photometry are available.  Of these, 298
stars have at least some near-IR photometry, while 287 have a complete
set of  {\it J-}, {\it H-}, and {\it K-} band measurements (with 280 of
these from the \tmass\/ program).  Table \ref{tabSTARS} lists all the
survey stars, along with some general descriptive information.  (The
complete version of Table \ref{tabSTARS} appears in the electronic
version of the paper.)  The stars in Table 1 are ordered by Right
Ascension.  For the star names, we adopted the most common form among
all the possibilities listed in the \simbad\/ database (i.e., ``HDnnn''
was the most preferred, followed by ``BDnnn'', etc.)  There are 185
survey stars which are members of open clusters or associations.  The
identity of the cluster or association is either contained in the star
name itself (e.g., NGC 457 Pesch 34) or is given in parentheses after
the star's name.  The spectral types in Table \ref{tabSTARS} were
selected from those given in the \simbad\/ database, and the source of
the adopted types is shown in the ``Reference'' column of the table.
When multiple types were available for a particular star, we selected
one based on our own preferred ranking of the sources.  For the B
stars, the quality of the spectral types varies widely, and the types
themselves are given only as a general reference --- {\it they do not
play any role in our analysis}.  A scan of the types seems to indicate
that in some instances we have violated our selection criteria, e.g.,
several ``Ib''s and ``e''s can be found.  However, in our estimation,
these are unreliable types and do not reflect the spectral information
we have examined.  For example, a number of B stars in clusters have
been erroneously classified as emission-line stars based on
contamination of their spectra by nebular emission lines.  The O stars
in the sample are, on the other hand, uniformly well-classified and the
types are used in the analysis (see \S \ref{secATLAS}).

An overview of the survey sample can be gained from Figures
\ref{figCOORDS} and \ref{figSTARSTATS}.  The former shows the location
of the stars on the sky (in Galactic coordinates) and the latter
summarizes the breadth of the stellar and interstellar properties of
the sample.  Figure \ref{figCOORDS} shows that our sample is clearly
biased to sightlines passing through the galactic plane, as would be
expected given by the lower limit of $\ebv = 0.20$ mag which we
imposed.  The locations of open clusters or associations for which
five of more members are in the survey sample are indicated on the
figure by the large circles.  Note that the size of the circles is {\it
not} intended to represent the physical extent of the
clusters or associations.  The data shown in histogram form in Figure
\ref{figSTARSTATS} are final results from the analysis, but are useful
here to characterize the sample.  Most of our stars are mid-to-early B
stars ($\teff = 15000-30000$ K) with a median reddening of
$\ebv = 0.45$ mag.  The median value of $R(V)$ ($\equiv
A_V/\ebv$) for the sample is 3.05, essentially identical to the Galactic
mean value for the diffuse ISM, and our sample is dominated by
sightlines though the diffuse ISM.  However --- due to the relatively
small sample size and the biases present in the \iue\/ satellite's
choice of targets over the years --- our survey does not necessarily
constitute a representative sample of the various types of regions
present in the ISM.  Care must be taken in interpreting average
properties derived from our results.  


\section{THE ANALYSIS\label{secMETHOD}}

It was shown by Paper IV that the energy distributions of reddened B-type
stars could be modeled successfully using theoretical predictions of
the intrinsic SEDs of the stars and a parametrized form of the
UV-through-IR extinction curve to account for the distortions
introduced by interstellar extinction.  A byproduct of the fit is a
determination of the wavelength-dependence of the extinction affecting
the star.  This is the essence of the extinction-without-standards
technique.  Although it was discussed in detail by Paper IV, for
completeness, we use this section to outline the basics of the
technique.  In addition, some details of the process have changed since
Paper IV, as a result of experience gained from the application of the
process to several hundred reddened stars.  These changes are
highlighted here.

\subsection{Modeling the SEDs}

The observed SED $f_{\lambda}$ of a reddened star can be represented as
\begin{equation}
\label{eqnFLUX}
f_{\lambda} = F_{\lambda}\, \theta_R^2\, 10^{-0.4 E(B-V) [k(\lambda-V) + R(V)]} \;\;,
\end{equation}
where $F_{\lambda}$ is the intrinsic stellar surface flux, $\theta_R
\equiv R/d$ is the stellar angular radius (where $R$ is the physical
radius and $d$ is the distance), \ebv\/ is the familiar measure of the
amount of interstellar reddening, $k(\lambda-V) \equiv
E(\lambda-V)/E(B-V)$ is the normalized extinction curve, and $R(V)
\equiv A_\lambda/E(B-V)$ is the ratio of reddening to extinction at
$V$.  By adopting stellar atmosphere models to represent $F_\lambda$
and a using parametrized form of $k(\lambda-V)$, we can treat Equation
(\ref{eqnFLUX}) as a non-linear least squares problem and solve for the
set of optimal parameters which generate the best fit to the observed
flux.  As in  Paper IV, we perform the least squares minimization using the
Interactive Data Language (IDL) procedure MPFIT developed by Craig
Markwardt \footnotemark \footnotetext{Markwardt IDL Library available
at http://astrog.physics.wisc.edu/$\sim$craigm/idl/idl.html.}.  The
observed SEDs which are fitted in the process consist of the \iue\/ UV
spectrophotometric fluxes, optical {\it UBV} magnitudes, and near-IR
{\it JHK} magnitudes discussed \S \ref{secDATA}.

As related in Paper IV, we developed this analysis utilizing R.L. Kurucz's
(1991) line-blanketed, hydrostatic, LTE, plane-parallel \atlas\/ models,
computed in units of erg~cm$^{-2}$~sec$^{-1}$~\AA$^{-1}$ and the
synthetic photometry derived from the models by Fitzpatrick \& Massa
(2005b).  The models are functions of four parameters: \teff\/,
\logg\/, [m/H], and $v_t$.  All of these parameters can be determined
in the fitting process although, because of data quality, it is
sometimes necessary to constrain one or more to a reasonable value and
solve for the others.  We smooth and bin the \iue\/ fluxes to match the
sampling of the \atlas\/ models (10 \AA\/ bins over most of the \iue\/
range; see Fitzpatrick \& Massa 1999).

Because we include some O stars in the current sample, we have expanded
our technique to incorporate the \tlusty\/ OSTAR2002 grid of
line-blanketed, hydrostatic, NLTE, plane-parallel models by Lanz \&
Hubeny (2003).  That grid includes 12 \teff\/ values in the range 27500
-- 55000 K, 10 chemical compositions from twice-solar to metal-free,
and surface gravities ranging from \logg\/ = 4.75 down to the modified
Eddington limit.  All models were computed with \vturb\/ = 10 \kms.  We
only consider the solar abundance models in this analysis and, thus,
the \tlusty\/ models are considered as functions of two parameters,
\teff\/ and \logg.  Synthetic {\it UBV} photometry for these models was
produced as described above for the \atlas\/ models.  To keep the
O-star and B-star fitting procedures as similar as possible, the
\tlusty\/ models were binned to the same wavelength scale as the
\atlas\/ models.  While analyzing the O stars we found that their
low-dispersion \iue\/ spectra (excluding the strong wind lines) are
very insensitive to temperature and our analysis yielded very uncertain
results.  As a result, we modified the procedure for these stars and
adopted values of \teff\/ based on their spectral types, rather than
solving for \teff.  Table \ref{tabOTEMPS} lists the temperature scale
used; it is a compromise between the results of Martins, Schaerer, \&
Hillier (2005) and our own analysis of optical line spectra of O-stars
(such as published by Walborn \& Fitzpatrick 1990) using the \tlusty\/
models, which are appropriate for O stars without massive winds.  The
results of this investigation will be reported elsewhere.  We assume a
generous uncertainty in these \teff\/ values (see below) and the
extinction curve results are actually very insensitive to the adopted
temperatures.

The value of the surface gravity $\log g$ is often poorly-determined
when using only broadband photometry, because it lacks a specific
gravity-sensitive index to help constrain $\log g$.  For the
cluster stars, however, we can apply ancillary information ---
specifically, the cluster distances as listed in Table \ref{tabSTARS}
--- to provide strong constraints on $\log g$.  We adopt the same
procedure as used by Fitzpatrick \& Massa (2005b), in which the Padova
grid of stellar structure models allow the Newtonian gravity of a star
to be inferred through its unique relation with the star's surface
temperature and radius.  (When the distance is specified, the physical
radius of the star becomes a fit parameter via its influence on the
angular radius $\theta_R$. Fitzpatrick and Massa (2005b) used distances
determined by \hip\/ parallaxes.)  In our iterative fitting procedure,
the current values of $T_{eff}$ and $\theta_R$, coupled with the Padova
models, determine the current value of $\log g$.  Generous 1-$\sigma$
uncertainties in the distances are included in the error analysis (see
below).  For field stars, no such constraints on $\log g$ are possible
and we solve for $\log g$ as a free parameter.  We do, however, apply a
reality check to the results and, if the final value seems physically
unlikely (i.e., $\log g >$ $\sim$4.3 or $\log g <$ $\sim$3.0), we
replace it with the mean sample value of  $\log g = 3.9$.
Uncertainties in this assumed value are incorporated in the error
analysis.

A parametrized representation of the extinction curve, covering the
whole UV-through-IR spectral range, is the heart of the current
analysis.  As in Paper IV, we construct this curve in two parts, joined
together at 2700 \AA.  An example of this formulation, for one
particular set of parameters,  is illustrated in Figure
\ref{figEXTCRV}.  In the UV, ($\lambda \leq 2700$ \AA; the shaded
region in the figure), we use a modified form of the UV parametrization
scheme of Fitzpatrick \& Massa (1990; hereafter Paper III), as shown by the
thick solid curve.  At longer wavelengths, we use a cubic spline
interpolation through a set of UV, optical, and IR ``anchor points''
(the $U_n$, $O_n$, and $I_n$ in the figure), as shown by the thick
dashed curve.

The modified Paper III extinction curve is defined by
\begin{eqnarray}
\label{eqnFMFUNC}
k(\lambda-V) = 
\left\{ 
\begin{array}{ll}
c_1 + c_2 x + c_3 D(x,x_0,\gamma)                  &  x \leq c_5     \\
c_1 + c_2 x + c_3 D(x,x_0,\gamma) + c_4 (x-c_5)^2  &  x > c_5  \;\; ,
\end{array}
\right. 
\end{eqnarray}
where $x \equiv \lambda^{-1}$, in units of inverse microns
(\invmic).  There are seven free parameters in the formula which
correspond to three features in the curve: (1) a linear component
underlying the entire UV wavelength range, defined by $c_1$ and $c_2$; (2)
a Lorentzian-like 2175 \AA\/ bump, defined by $c_3$, $x_0$, and $\gamma$
and expressed as
\begin{equation}
\label{eqnDRUDE}
D(x,x_0,\gamma) = \frac{x^2}{(x^2-x_0^2)^2 +x^2\gamma^2} \;\; ;
\end{equation}
and (3) a far-UV curvature component (i.e., the departure in the far-UV
from the extrapolated bump-plus-linear components), defined by $c_4$
and $c_5$.  All seven free parameters can be determined by the least
squares minimization algorithm.

The modification made here to the Paper III formula is in the far-UV
curvature term.  In Paper III, the value of $c_5$ was fixed at 5.9 \invmic\/
and $c_4$ was the scale factor applied to a pre-defined cubic
polynomial.  In working with the current dataset, we found that the
formulation in Equation (\ref{eqnFMFUNC}) significantly improved the fits to
many stars, particularly those with weak far-UV curvature, and degraded
the fits in almost no cases.  In the modified form, the curvature is
functionally simpler --- containing a single quadratic term ---
although we have added another free parameter to the extinction
curve representation.  Because the primary goal of the Paper III formula was
(and still is) to provide an analytical expression which reproduces as
closely as possible observed extinction curves, the cost of an
additional free parameter was deemed worthwhile.

Using the UV fit parameters above, additional quantities can be defined
which help describe the UV curve properties.  Particularly useful ones
include 1) $\Delta1250 \equiv c_4(8.0-c_5)^2$, which is the value of
the FUV curvature term at 1250 \AA\/ and provides a measure of the
strength of the FUV curvature; 2) $A_{bump} \equiv \pi \;
c_3/(2\gamma)$, which is the area of the 2175 \AA\/ bump; and 3)
$E_{bump} \equiv c_3/\gamma^2$, which is the maximum height of the
2175 \AA\/ bump above the background linear extinction.

The cubic spline interpolation which produces the optical-through-IR
region of our parametrized extinction curve is produced using the IDL
procedure SPLINE.  The nine anchor points shown in Figure
\ref{figEXTCRV} are specified by five free parameters.  The UV points,
$U_1$ and $U_2$, are simply the values at 2600 \AA\/ and 2700 \AA\/
resulting from the modified Paper III formula and require no new free
parameters.  They assure that the two separate pieces of the extinction
curve will join smoothly, although not formally continuously.  The
three optical points, $O_1$, $O_2$, and $O_3$ (at 3300 \AA, 4000 \AA,
and 5530 \AA, respectively) are each treated as free parameters and are
adjusted in the fitting procedure to assure the normalization of the
final extinction curve.  The IR points, $I_1$, $I_2$, $I_3$, $I_4$, and
$I_5$ (at 0.0, 0.25, 0.50, 0.75, and 1.0 \invmic, respectively) are
functions of two free parameters, $k_{IR}$ and $R(V)$ as follows:
\begin{equation}
I_n \equiv k(\lambda - V) = k_{IR} \lambda_n^{-1.84} - R(V) \;\; .
\label{eqnIR}
\end{equation}
This assures that the IR portion of our curve follows the the power-law
form usually attributed to IR extinction, with a value for its exponent
from Martin \& Whittet (1990).  As noted in Paper IV, the value of the
power law exponent could potentially be determined from an analysis
like ours.  However, we have found that an IR dataset consisting only
of {\it JHK} magnitudes, as in our survey, is insufficient to specify
three IR parameters.

Note that we adopt the cubic spline formulation for the optical/IR
extinction curve simply because we do not have an acceptable analytical
expression for the curve shape over this range.  The spline approach is
very flexible in that the number of anchor points can be modified
depending on the datasets available.  In the current application, we
use only three optical anchor points because we have only three optical
data points ($V$, $B-V$, and $U-B$).  This approach will bear its best
fruit when it can be applied to spectrophotometric data in the near-IR
through near-UV region.  Then, a large number of anchor points can be
used to precisely measure the heretofore poorly-determined shape of
extinction in this region, without bias towards a particular analytical
expression.

In summary, the analysis performed here --- modeling the SEDs of 328
reddened stars via Equation (\ref{eqnFLUX}) --- involves determining the
best-fit values for as many as 18 free parameters per star, via a
non-linear least squares analysis.  These include up to 4 parameters to
define the theoretical stellar atmosphere model (\teff, \logg, \abund,
\vturb), up to 12 parameters to describe the extinction curve shape
($O_1$, $O_2$, $O_3$, $R(V)$, $k_{IR}$, $c_1$ through $c_5$, $x_0$,
$\gamma$), the angular radius $\theta_R$, and \ebv.  We weight the UV,
optical, and IR datasets equally in the fitting procedure.

\subsection{Error Analysis
\label{secERRORS}}
 
One of the main benefits of the extinction-without-standards technique
is the error analysis, which provides a well-quantified estimate of the
uncertainties in the best-fit model parameters and allows possible
correlations between parameter errors to be explored.  This latter
benefit is important for assessing the reality of apparent correlations
between parameters.  The uncertainties in the best-fit parameters
derived for our survey stars were determined by running 100 Monte Carlo
simulations for each star.  In each simulation, the fitting procedure
was applied to an input SED consisting of the final best-fit model
convolved with a random realization of the observational errors
expected to affect the actual data.  The adopted 1-$\sigma$
uncertainties for each parameter, which will be presented in \S
\ref{secATLAS}, were taken as the standard deviations of the values
produced by the 100 simulations.

Our observational error model for the \iue\/ data consists of random
photometric uncertainties and camera zero-point errors as described by
Fitzpatrick \& Massa (2005b).  The assumed observational errors in the
Johnson $B-V$ and $U-B$ indices were as given in Table 7 of that
paper.  The $V$ magnitudes were assumed to have a 1-$\sigma$
uncertainty of 0.015 mag.  The uncertainties in the \tmass\/ $JHK$ data
were as obtained from the \tmass\/ archive.  Johnson $JHK$ magnitudes
were assumed to have uncertainties of $\pm$0.03 mag.  Random
realizations of each of these observational errors, which were added to
the best-fit model SED for each Monte Carlo simulation, were determined
using the IDL procedure RANDOMN, which produces a normally distributed
random variable.
  
In cases where assumptions were made about the values of specific fit
parameters, we incorporated uncertainties in the assumptions in the
error analysis.  In particular: (1) for the O-type stars, the adopted
spectral type-dependent temperatures were taken to have 1-$\sigma$
uncertainties of $\sim$1000 K; (2) for cluster stars, the adopted
distances (used to constrain the surface gravities) were assumed to
have 1-$\sigma$ uncertainties of $\pm$20\%; and 3) for the field stars
whose values of $\log g$ were taken to be the sample mean of 3.9, this
mean was assumed to have a 1-$\sigma$ uncertainty of $\pm$0.2.  The
values of $\teff$, distance, and $\log g$ used in the the Monte Carlo
calculations for the relevant stars were varied randomly in the
simulations (using RANDOMN) in accord with these uncertainties.

This study certainly does not constitute the first attempt to quantify
the uncertainties in interstellar extinction curves.  Most pair method
studies (see, for example, Cardelli, Sembach, \& Mathis 1992) have
incorporated some form of error analysis, often based on the
methodologies presented by Massa, Savage, \& Fitzpatrick (1983) and
Massa \& Fitzpatrick (1986).  However, as long as we restrict our
sample to stars which are well represented by the model atmospheres we
employ, the advantages of the current technique are great.  Because the
stellar parameters (temperature, surface gravity, and abundance) are
given by continuous mathematical variables (instead of a non-uniformly
sampled, discrete sets of standard stars), we are able to perform a
well-defined Monte Carlo analysis.  The results of this analysis
explicitly quantify the uncertainties in all of the input data and
assumptions and, thus, the final error bars affecting the derived
curves.  Moreover, since many realizations of the individual curves are
produced, the full shape of the ``error ellipses'' (which describe
correlations between the errors) are determined for each specific set
of input parameters.  Additional discussion of the
extinction-without-standards error analysis can be found in Paper IV of
this series, along with a demonstration of the quantitative accuracy of
the results.


\section{AN ATLAS OF GALACTIC EXTINCTION CURVES\label{secATLAS}}

The results of the extinction-without-standards analysis of 328
Galactic stars are presented in Table \ref{tabPARMS}, Table
\ref{tabEXTINCT}, and Figure \ref{figATLAS}.  The 18 free parameters
determined by the fitting procedure are divided between Tables
\ref{tabPARMS} and \ref{tabEXTINCT}, with the latter containing the 12
parameters which define the shape of the normalized interstellar
extinction curves $k(\lambda-V) \equiv E(\lambda-V)/E(B-V)$.  For both
tables, only the first 10 entries are shown here.  The full versions of
the tables can be viewed in the electronic edition of the Journal.  The
uncertainties listed in the tables are the 1-$\sigma$ errors derived
from the Monte Carlo analysis described in \S \ref{secERRORS}.

Figure \ref{figATLAS} shows the normalized extinction curves for the
survey sample.  The figure consists of 33 panels, each (except the
last) containing 10 extinction curves arbitrarily shifted vertically
for clarity. Only the first panel is shown here.  The full figure is
given  in the electronic version of this paper.  The solid curves in
the figure show the parametrized UV-through-IR curves whose shapes were
determined by the fitting procedure described in \S \ref{secMETHOD}
(the parameters describing the curves are in Table \ref{tabEXTINCT}).
An estimate of the shape of the average Galactic extinction curve
(corresponding to $R(V)=3.1$; from Fitzpatrick 1999) is shown for
reference by the dash-dot curves.  The 1-$\sigma$ uncertainties of the
survey extinction curves are indicated in Figure~\ref{figATLAS} by the
grey shaded regions, which are based on the Monte Carlo error
simulations.  Their thicknesses indicate the standard deviations of the
ensemble of simulations at each wavelength.  The actual normalized
ratios between the observed stellar SEDs and the atmosphere models are
shown by the symbols.  Large filled circles indicate {\it JHK} data in
the IR region ($\lambda^{-1} < 1 \; \invmic$) and {\it UBV} data in the
optical region ($1.5 < \lambda^{-1} < 3.0 \; \invmic$).  In the UV
region (($\lambda^{-1} > 3.3 \; \invmic$), the small symbols with
1-$\sigma$ error bars show the ratios between the \iue\/ data and the
models.

A close examination of the curves in Figure \ref{figATLAS} shows that
the parametrized curves are extremely good representations of the
observed extinction ratios and thus serve as useful proxies for the
actual curves themselves.  This is particularly apparent in the UV,
where the spectrophotometric data show the flexibility of the
parametrization scheme.  For those wishing to use these curves in
extinction studies, we have prepared a tar file containing the
parametrized curves for all 328 stars, sampled at 0.087 \invmic\/
intervals, and their accompanying fit parameters.  Directions for
retrieving the file are given in the Appendix.

In using or interpreting these curves it is important to recognize that
their shapes in the regions between the IR and optical and between the
UV and optical are interpolations only and not strongly constrained by
data.  Additional observations, particularly fully-calibrated
spectrophotometric data, would be very useful to constrain the shape of
the extinction curves in these regions --- and in the optical where
only broadband {\it UBV} measurements are currently employed.  It is
certainly counter-intuitive that the spectral regions where the
detailed shapes of the extinction curves are most poorly-determined are
ground-accessible, while the UV data are so well-measured.


\section{Properties of the Sample Stars\label{secSTARS}}

Although the main goal of this paper is to explore Galactic extinction,
it is nevertheless reasonable to consider briefly the stellar
properties revealed by our analysis since they directly affect the
extinction results.  As discussed above, our reliance on broadband
photometry for the current work results in stellar parameters that are
not as accurate as those presented in Paper IV, due to the lack of a
good surface gravity discriminator.  Nevertheless, more than 50\% of
the sample stars reside in open clusters and associations and, for
these stars, the accuracy will be increased, due the use of ancillary
information.

Of the four stellar properties determined in the analysis (\teff,
\logg, \abund, \vturb), the most significant to our extinction program
is \teff\/ because it has the most impact on the shapes of the model
atmosphere SEDs used to derive the extinction curves.  Figure
\ref{figSPTEFF} shows a plot of the derived \teff\/ values against
spectral type for the spectral class B0 and later stars (filled and
open circles).  For comparison, we show several spectral type vs.
\teff\/ calibrations from the literature (solid and dashed lines) and
data from our photometric calibration study (Fitzpatrick \& Massa
2005b, open squares) in which we modeled the SEDs of 45 unreddened
B-type stars.  While the survey star data are generally consistent with
the comparison data in the figure, considerable scatter is present at
some spectral classes --- most particularly at types B2 and B3 --- and
there is a general departure between our results and the \teff\/
calibrations in the neighborhood of types B1 and B2, in the sense that
our results indicate hotter temperatures than the calibrations.
 
We examined the \iue\/ spectra of a number of the survey stars ---
those whose temperatures are most discrepant with their claimed
spectral types --- and compared them with spectral classification
standards to see if somehow our fitting procedure was arriving at
grossly incorrect temperatures.  An example of such a comparison is
shown in Figure \ref{figCOMPARE}.  We plot a portion of the \iue\/
spectra of the survey stars HD228969 and BD+45 973, the two hottest
stars in the ``B2'' and ``B3'' spectral bins, respectively, along with
several spectral standards with expected temperatures in the
neighborhood of 30000 K.  The close match between the survey stars and
the hot standards is evident and it is clear that the cool spectral
types found in the literature for these stars are unreliable.  This is
the general conclusion from all such comparisons we have performed.
The temperatures found from the fitting procedure are consistent with
those expected from a close examination of the UV spectral features of
the stars.  We conclude that the outliers in Figure \ref{figSPTEFF}
result from poor optical spectral types.  This is not surprising, since
the available types are from a large number of sources and based on a
wide variety of observational material of very non-uniform quality.

The general discrepancy between our results and the calibrations in the
B1-B2 region is a different matter.  We derive considerably hotter
effective temperatures for the B1 (25000 -- 26000 K) and B2 (23000 --
24000 K) stars than expected from previous calibrations.  However,
inspection of the UV features in the \atlas\/ models make it difficult
to believe that typical B1 and B2 stars are as cool as the spectral
type -- \teff\/ calibrations suggest.  We must bear in mind, however,
that our sample is strongly biased toward cluster stars, which may be
considerably younger and more compact than the ``field'' B stars used
in the calibrations.  Furthermore, the current B star calibrations are
all over 20 years old, and it is quite possible that they are in need
of revision.

Another way to look at the \teff\/ values is shown in Figure
\ref{figCBRACKET} where we plot \teff\/ as a function of the
Str\"{o}mgren reddening-free index $[c] \equiv c_1 - 0.20(b-y)$, which
is a measure of the strength of the Balmer jump.  Str\"{o}mgren
photometry is not used in this program but is available for 162 of our
stars.  The symbols in Figure \ref{figCBRACKET} are the same as in
Figure \ref{figSPTEFF}, with the addition of the open circles which
denote O stars.  The figure demonstrates the essentially exact overlap
between the current results and those for the unreddened, mid-to-late B
stars from Fitzpatrick \& Massa (2005b) as well as the smooth
transition from the early B stars into the O stars.  There is no
indication of any systematic effects present in the results for the
early B stars.  On the contrary, the spectral type vs. \teff\/
relations, transformed into the \teff\/ vs. $[c]$ plane as described in
the figure legend, show a number of abrupt and physically unrealistic
changes in slope suggestive of inadequacies in the calibrations.

We conclude that our derived effective temperatures are reasonable.  In
most cases where a temperature strongly disagrees with a published MK
type, it agrees quite well with the UV type determined by Valencic,
Clayton, \& Gordon (2004), indicating that the MK type is of poor
quality or else influenced by something else (e.g., the presence of a
cooler companion which is invisible in the UV).  We also suspect that
the spectral type -- \teff\/ calibration for the B1 and B2 stars may
need to be revised.  Finally, we are gratified by the overall
consistency between the current results and those of Paper~IV and
Fitzpatrick \& Massa (2005b), where the stellar parameters were more
strongly constrained, and with the smooth relation between \teff\/ and
[c], which suggests that no strong systematic effects are present.  


\section{Properties of Galactic Interstellar Extinction\label{secEXTINCTION}}

\subsection{General Properties\label{secEXT_PROPERTIES}}

Although the individual extinction curves for all of the survey stars
are displayed in the 33 panels of Figure \ref {figATLAS}, it is
nonetheless difficult --- from that figure --- to visualize the range
of extinction properties present in the sample.  To provide such a
view, we plot the analytical fits for the full set of 328 survey curves
in Figure \ref{figALLCURVES}.  (These curves can be reproduced from the
parameters listed in Table \ref{tabEXTINCT}.  The curves have been
plotted using small dots in those spectral ranges where they are
interpolated or extrapolated.  The solid portions correspond to the
regions constrained by near-IR {\it JHK} photometry, optical {\it UBV}
photometry, and UV \iue\/ spectrophotometry.  The top panel, in which
the curves are plotted in their native form, $E(\lambda-V)/E(B-V)$,
shows the wide range of variation observed in Galactic extinction
curves, although a clear core of much more restricted variation is
evident in the distribution.  The convergence of the curves in the
range $1.8 < \lambda^{-1} < 2.5~\invmic$ is a product of the
normalization and obscures our view of variations in the optical
region.  The bottom panel of Figure \ref{figALLCURVES} presents the
same curves, but normalized to the total extinction at 1~\invmic.  The
exact overlap of the curves at $\lambda^{-1} < 1~\invmic$ arises
because we have adopted a power law form with a fixed exponent of
$-1.84$ for all curves in this spectral region (see Eq. [\ref{eqnIR}]).
It is unlikely that the exponent is actually so constant --- Larson \&
Whittet (2005) have found evidence for a more negative value in a
sample of high-latitude clouds --- however, as noted earlier in \S
\ref{secMETHOD}, our near-IR dataset consisting of only {\it JHK}
measurements is insufficient for independently evaluating the
exponent.  This IR normalization reveals that the extinction in the
optical spectral region can range from very-nearly grey to very
strongly wavelength-dependent.

One of the goals of many Galactic extinction studies is to derive an
estimate of the typical or average wavelength-dependence of
interstellar extinction.  Such a mean curve is often used as the
standard of normalcy against which particular sightlines are judged, or
for comparison with results for external galaxies, or for
``dereddening'' the SEDs of objects for which there is no specific
extinction knowledge.  While constructing an average curve is a
straightforward process, the degree to which the result represents
``normal'' or even ``typical'' extinction is problematic.  We will take
up the issue of an average Galactic curve later in \S
\ref{secDISCUSS}.  Here we present an average curve for our sample, to
help characterize the general extinction properties of our survey
sightlines.

The bottom panel in Figure \ref{figSTARSTATS} shows a clear peak in the
distribution of $R(V)$ values of our sample.  A Gaussian fit to the
region of this peak, shown as the smooth curve in the figure, has a
centroid at $R(V) = 2.99$ and a width given by $\sigma = 0.27$.  The
peak is within the range of values considered as average for the
diffuse ISM (see, e.g., Savage \& Mathis 1979).  Thus, as noted in \S
\ref{secDATA}, our sample is dominated by sightlines whose $R(V)$
values are consistent with the diffuse phase of the ISM (although
composite sightlines are undoubtedly present).  We have constructed an
average curve to represent the properties of these diffuse sightlines
by taking the simple mean extinction value at each wavelength using all
the curves with $2.4 < R(V) < 3.6$ (i.e., the ~2-$\sigma$ range of the
Gaussian fit in Figure \ref{figSTARSTATS}; 243 sightlines in total).
This mean extinction curve is shown in Figure \ref{figAVGCURVE} by the
thick solid curve.  The dark grey shaded region shows the variance of
the 243-curve sample.  The set of 12 parameters describing this curve
are listed in Table \ref{tabAVGCURVE}.  Since so many of our sightlines
are included in the mean, removing the restriction on $R(V)$ has little
effect.  If we had included all 298 sightlines for which $R(V)$ has
been derived, then the mean curve would differ from that shown only by
being several tenths lower in the UV region.  The variance of the full
sample is larger, and this is shown by the lightly shaded region.  For
comparison, several other estimates of average Galactic curves are
shown in Figure \ref{figAVGCURVE}.  The curves from Cardelli et al.
(1989), Fitzpatrick (1999), Seaton (1979), and Valencic et al. (2004) are all intended to represent the diffuse ISM mean.  The
results from Savage et al. (1985) are mean values for the 800
sightlines in that study with $E(B-V) \ge 0.20$ mag (matching our
survey cutoff).  No restriction                                                                                                   on $R(V)$ was imposed.  The error bars for
the Savage et al. (1985) data are sample variances; they are generally
similar to the variances of our full sample (lightly shaded region).
The much larger value for the 1500 \AA\/ point is likely due to
spectral mismatch in the potentially strong C IV stellar wind lines.

The differences among the various mean curves in Figure
\ref{figAVGCURVE} are instructive.  The great intrinsic variety of
Galactic extinction curves as seen, for example, in Figure
\ref{figALLCURVES} shows that any mean curve is subject to the biases
in the sample from which it was produced.  It is probably impossible to
construct a sample of sightlines whose properties could be claimed to
provide a fair representation of all the types of conditions found in
the ISM.  Thus, there is likely no unique or best estimate of mean
Galactic extinction.  Any of the mean curves in Figure
\ref{figAVGCURVE} would serve as reasonable representations of Galactic
extinction.  In any situation where an average curve is adopted,
however, it is important to recognize the intrinsic variance of the
underlying sample and incorporate the uncertainties of the average
curve in any error analysis.  Because it is derived from such a large
sample and is largely free of contributions from spectral mismatch, the
sample variance for our diffuse curves (shown in Figure
\ref{figAVGCURVE} by dark shaded region) would provide a reasonable
estimate of the uncertainty in any version of a mean Galactic diffuse
ISM curve.  We have included our diffuse ISM mean curve from Figure
\ref{figAVGCURVE} --- and its accompanying uncertainty --- in the tar
file discussed in \S \ref{secATLAS} (see the Appendix).

We can also use our large sample to investigate the ``smoothness'' of
UV extinction.  In Figure \ref{figIUECURVE}, we plot the simple mean of
the actual extinction ratios for our survey sample in the spectral
region covered by the \iue\/ data (upper curve, open circles).
Overplotted is a parameterized fit to these data, using the extinction
formulation given in \S \ref{secMETHOD} (solid curve).  This figure
illustrates two points; namely, 1) the lack of small-scale structure in
UV extinction and 2) the degree to which the modified Paper III UV
parametrization scheme reproduces the shape of UV extinction.  A
detailed discussion of the former point, and an indication of the kinds
of features that might be expected in the UV, can be found in Clayton
et al. (2003).  In short, polycyclic aromatic hydrocarbon (PAH)
molecules, which have been suggested as the source of mid-IR emission
features, might produce noticeable absorption in the UV, and possibly
even contribute to the 2175 \AA\/ bump absorption.  Earlier studies
have always failed to find structure in UV extinction curves and
Clayton et al. were able to place very stringent 3-$\sigma$ upper
limits of $\sim0.02A_V$ on any possible 20 \AA-wide features in the
extinction curves towards two heavily reddened B stars.  The data at
the bottom of Figure \ref{figIUECURVE} show the differences between the
mean extinction curve and the best-fit model.  A small number of points
do rise above (or below) the general noise level of the residuals, but
these points --- which are labeled in the figure --- are due to
interstellar gas absorption lines (which are sometimes strong in the
observations and always non-existent in the model atmosphere SEDs),
mismatch between the C IV $\lambda$1550 stellar wind lines in the O
stars and the static model SEDs, or known inadequacies in the \atlas\/
opacity distribution functions (labeled ``b'' in the figure; see Paper
IV).  Excluding these points, the standard deviation of the mean curve
around the best-fit is 0.06\ebv\/ mag, corresponding to $\sim0.02A_V$
mag, for the 10 \AA-wide spectral bins.  This is not quite so
restrictive as the result of Clayton et al., but once again affirms the
smoothness of UV extinction.

\subsection{Spatial Trends \label{secEXT_SPATIAL}}

Studying spatial trends in extinction can serve two purposes.  First,
identifying strong regional variations can allow a better estimate of
the shape of the extinction curve affecting a particular sightline
permitting, for example, a more accurate determination of the intrinsic
SED shape of an exotic object.  This has been the motivation for a
number of studies, such as by Massa \& Savage (1981) and Torres (1987)
who used extinction curves derived for B stars in open clusters (NGC
2244 and NGC 6530, respectively) to determine the SEDs of cluster O
stars.  The second purpose is to gain insight into the nature of dust
grains and the processes which modify them by observing how extinction
curves respond to various environmental properties, such as density or
radiation field.

In Figure \ref{figLONGITUDE} we show a sweeping view of the regional
trends in our data by plotting several extinction curves properties
against Galactic longitude for each of our survey sightlines.
Sightlines towards stars in the clusters or associations highlighted in
Figure \ref{figCOORDS} are indicated by the larger symbols in Figure
\ref{figLONGITUDE} (the key is in figure caption) and the rest of the
sample by the small filled circles.  The dashed lines in the figure
panels show the parameter values that correspond to the diffuse mean
curve in Figure \ref{figAVGCURVE}.

A number of regions stand out in Figure \ref{figLONGITUDE}.  For
example, NGC 1977 near $l = 210\degr$ (open triangles; includes the Orion
Trapezium region) is well-known for its high $R(V)$ sightlines.  Figure
\ref{figLONGITUDE} shows that this region is also notable for its flat
UV extinction curves (i.e., small $c_2$) and weak bumps (i.e., small
$A_{bump}$).  Interestingly, the far-UV curvature for this region
appears typical.  An early discussion of UV extinction curves towards
NGC 1977 can be found in Panek (1983).  The relationship between $R(V)$
and other curve properties will be the subject of \S
\ref{secEXT_RDEPEND} below.  

The Carina direction (large filled circles), particularly towards Tr 14
and Tr 16, also shows elevated $R(V)$ values.  This is another
weak-bumped direction which also shows lower than average FUV
curvature.  Optical and IR extinction studies have been performed for
Carina sightlines (e.g., Tapia et al. 1988 and references within), but
we are not aware of correspondingly detailed UV extinction studies.
  
Large $R(V)$ values are also seen along a number of sightlines in the
general direction of the Galactic center.  This includes sightlines to
the cluster NGC 6530 (open squares) and towards the $\rho$ Oph dark
cloud (small filled circles near $l \simeq 253 \degr$.  These
sightlines also feature low UV extinction and, in the case of NGC 6530,
weaker than average 2175 \AA\/ bumps.  Large $R(V)$ are well known in
the Ophiuchus region (Chini \& Krugel 1983) and the UV extinction has
been examined by Wu, Gilra, \& van Duinen (1980).  UV extinction
towards NGC 6530 was studied by Torres (1987).

Finally we note a broad region from $l \simeq 50 \degr$ to $l \simeq 150
\degr$ where the $R(V)$ values are systematically slightly below the mean
value.  No trends are obvious in the UV parameters.  These sightlines
are in the direction of the Perseus spiral arm (e.g., Georgelin \&
Georgelin 1976) and sample dust in the interarm region and, possibly,
the Perseus Arm itself --- for those stars more distant than $\sim$2
kpc, such as in h \& $\chi$ Per (x's in Figure \ref{figLONGITUDE}).
Extinction towards individual regions in this zone have been studied
(e.g., Tr 37 by Clayton \& Fitzpatrick 1987, Cep OB3 by Massa \& Savage
1984, h \& $\chi$ Per by Morgan, McLachlan, \& Nandy 1982, and Cr 457
by Rosenzweig \& Morrison 1986), but we are not aware of any
comprehensive investigation of the general region.  

Morgan et al. (1982) found a dependence of UV extinction on Galactic
latitude, $b$, for sightlines to stars in h \& $\chi$ Per, in the sense
that the extinction at 1550 \AA\/ increases with increasingly negative
values of $b$.  Our data suggest a similar effect for the UV linear
slope $c_2$ (which is closely related to the extinction level at 1550
\AA), but not for any other extinction parameter, including $R(V)$.  We
have examined our dataset for general trends with Galactic latitude, or
with distance above and below the plane, and have found none.  This is
not surprising, however, since our sample is dominated by low-latitude
sightlines and we have little leverage for a latitude-dependence
study.  Our lower cutoff of 0.20 mag in \ebv\/ eliminated most
high-latitude sightlines from consideration.  Previous studies have
shown that it is difficult to uncover latitude dependences in
extinction (e.g., Kiszkurno-Koziej \& Lequeux 1987).  Local trends
might be uncovered by examining small zones in Galactic longitude (such
as in the study by Morgan et al.), although such detailed
investigations are beyond the scope of this paper.  Likewise, studies
of extinction variations over small spatial scales, such as among
sightlines to cluster members are beyond our scope, but might well
provide important information linking grain populations with other ISM
diagnostics.

The quantity $\ebv/d$, which can be computed from the data in Tables
\ref{tabSTARS} and \ref{tabPARMS}, provides a crude measure of one
important physical property of the ISM, namely, dust density.  While
the shortcomings of this measure as a direct proxy for density are
clear --- for example, a high density dust cloud along a long,
otherwise vacant sightline will yield a misleadingly low value of
$\ebv/d$ --- it is nevertheless useful as a first-look and as a guide for
future studies.  Figure \ref{figDENSITY} shows plots of four extinction
curve parameters against $\ebv/d$.  The symbols are the same as for
Figure \ref{figLONGITUDE}.  The three UV parameters all show evidence
for a weak trend with density, in the sense of flatter slopes, broader
bumps, and increasing FUV curvature with increasing density.  Hints of
these three effects were seen in the first two papers in this series
(Fitzpatrick \& Massa 1986, 1988; hereafter Papers I and II).  We see
no evidence for a trend with bump strength $A_{bump}$ (not shown in
Figure \ref{figDENSITY}) and the dependence of $R(V)$ on $\ebv/d$ is
complex and difficult to characterize, although the highest density
sightlines all have larger-than-average values of $R(V)$.  While not
conclusive, the results in Figure \ref{figDENSITY} certainly suggest
that comparisons of our survey data with detailed measures of ISM
physical conditions could yield interesting results.  We note that
Rachford et al. (2002) have found positive trends of bump width and
far-UV curvature with the fraction of hydrogen atoms in the form
H$_2$.  This is consistent with the results in Figure \ref{figDENSITY}
in the sense that one would expect higher H$_2$ fractions in denser,
and therefore better-shielded, regions.  Other studies of the
density-dependence of extinction have been performed by Massa (1987)
and Clayton, Gordon, \& Wolff (2000).

\subsection{Relationships Among the Fit Parameters\label{secEXT_RELATIONS}}

With twelve parameters to describe the UV-through-IR extinction curves
of each of the 328 survey stars, there are many possible correlations
and relationships to investigate.  We have looked at all of these
possibilities and, as an interested reader can verify from the data in
Table \ref{tabEXTINCT}, virtually all the parameters are remarkably
UNcorrelated with each other!  In this section, we consider only the
two most striking relationships between extinction parameters: $c_1$
vs. $c_2$ (the UV linear intercept and slope) and $R(V)$ vs. $k_{IR}$
(the ratio of selective-to-total extinction and the IR scale
factor).  We also examine the most important non-relationship between
parameters: $x_0$ vs. $\gamma$ (the centroid and width of the 2175
\AA\/ bump).  Below, in \S \ref{secEXT_RDEPEND}, we will consider the
possible relationship between IR curve features and UV features.  
\subsubsection{$c_1$ vs. $c_2$}

In Papers I and II we
showed that pair method extinction curves in the region of the 2175 \AA\/
extinction bump could be modeled very precisely using a Lorentzian-like
``Drude profile'' (see Eq. [\ref{eqnDRUDE}]) combined with a linear
background extinction (defined by the parameters $c_1$ and $c_2$ in
Eq. [\ref{eqnFMFUNC}]).  Further, it was shown that the linear
parameters for the 45 sightlines in the study appeared to be very well
correlated, and could likely be replaced with a single parameter
without loss of accuracy.  In the pilot study for this paper, F04
showed that this correlation was maintained in a larger sample of 96
curves derived using the extinction-without-standards technique.

Figure \ref{figC1C2} shows a plot of the linear parameters $c_1$ vs.
$c_2$ for our survey sample of 328 extinction-without-standards
curves.  The dotted error bars show the orientation of the 1-$\sigma$
error ellipses of the measurements. These were determined by the
distribution of results from the Monte Carlo simulations.  The obvious
correlation between the errors in $c_1$ and $c_2$ is not unanticipated
(see Figure 4 in Paper II), but the ability to
explicitly determine such errors is a major advantage of the
extinction-without-standards approach and is critical for evaluating
the significance of apparent correlations.

The solid line in Figure \ref{figC1C2} corresponds to the linear relation
\begin{equation}
\label{eqnC1C2}
c_1 = 2.09 - 2.84 \; c_2 \;\; ,
\end{equation}
which is a weighted fit that minimizes the scatter in the direction
perpendicular to the fit.  (Because there is uncertainty in both $c_1$
and $c_2$, a normal least squares fit is not appropriate.)  The fact
that this relationship is nearly parallel to the long axis of the
correlated error bars explains why the relationship between $c_1$ and
$c_2$ remains so clear, even in the presence of observational error.  

To determine whether the observed scatter about the mean $c_1$ vs $c_2$
relationship is caused only by observational error or is at least
partially the result of ``cosmic'' scatter, we must examine the
residuals to the best-fit relationship.  The distribution of residuals
perpendicular to the best-fit line is complex, consisting of a Gaussian
core of values with $\sigma \simeq 0.07$ and a more extended
distribution of outliers reaching out to values of about $\pm0.3$. About 
86\% of the points fall within the 2-$\sigma$ range
of the Gaussian core.  The RMS value of the expected observational
errors perpendicular to the best-fit relation is 0.057.  The Gaussian
core of the observed residuals is thus only slightly broader than that
expected from observational scatter alone.  We conclude that $c_1$ and
$c_2$ are indeed intrinsically well-correlated quantities, with a
cosmic scatter comparable to our measurements errors.  However a
significant fraction of the sample ($\sim$10\%) show evidence for a
wider deviation from the mean relationship.  In most instances,
Equation (\ref{eqnFMFUNC}) could be simplified without loss of accuracy by
replacing the two linear parameters $c_1$ and $c_2$ with a single
parameter.

\subsubsection{$R(V)$ vs. $k_{IR}$}

F04 showed that the two parameters describing the IR portion of the
extinction-without-standards curves, $k_{IR}$ and $R(V)$ are apparently
well-correlated and that the IR curve might be defined by a single
parameter.  Figure \ref{figRKIR} shows a plot of $k_{IR}$ vs. $R(V)$
for the survey sample, along with the 1-$\sigma$ error bars.  The solid
curve shows the best-fit weighted linear relationship
\begin{equation}
\label{eqnKRV}
k_{IR} = -0.83 + 0.63 \; R(V) \;\; ,
\end{equation}
which minimizes the scatter in the direction perpendicular to the
relation.  As in the discussion above, we see that the errors are
strongly correlated and the near coincidence of the long axis of the
error ellipses with the direction of the best-fit relationship would
preserve the appearance of a correlation even in the face of
significant observational error.  The residuals in the direction
perpendicular to the best-fit line are distributed in a Gaussian form,
with $\sigma = 0.11$.  The RMS value of the observational errors in
this same direction is $\pm0.12$.  Thus the observed distribution of
points in Figure \ref{figRKIR} is consistent with perfectly correlated
quantities and the expected observational error.  

The correlation between $R(V)$ and $k_{IR}$ indicates that the shape of
near-IR extinction at $\lambda >$ $\sim$1 $\mu {\rm m}$, over a wide
range of $R(V)$ values, can be characterized by a single parameter.
I.e., the two parameters in Equation (\ref{eqnIR}) are redundant and
could be replaced --- without loss of accuracy --- by a single
parameter, e.g., $R(V$), based on Equation (\ref{eqnKRV}):
\begin{equation}
\label{eqnIREXT}
\frac{E(\lambda-V)}{E(B-V)} =  [-0.83+0.63R(V)]\lambda^{-1.84} - R(V)   \;\;   .
\end{equation}
This is consistent with the results of Martin \& Whittet (1990; see
their Table 2), who utilized IR data out to the $M$ band near 5 $\mu$m.
Our study utilizes only shorter wavelength near-IR bands, but includes
many more sightlines and spans a wider range in $R(V)$ values than
could be studied by Martin and Whittet.  Our parameter $k_{IR}$ is
essentially the same as their $e$ parameter, and our results
demonstrate the dependence of $e$ on $R(V)$.

The significance of the tight correlation between $R(V)$ and
$k_{IR}$ is somewhat difficult to assess.  We must keep in mind that,
at one level, this relation simply states that the three IR data points
given by \tmass\/ $JHK$ photometry can be summarized by two parameters
at the level of the errors in the \tmass\/ photometry.  On the other
hand, it also bears on the issue of the underlying shape of infrared
extinction and its so-called universality.  Our data, which are
dominated by diffuse ISM sightlines, are consistent with the notion of
a universal shape for IR extinction, as given by Equation
(\ref{eqnIREXT}).  However, given our restricted IR wavelength coverage
and the typical uncertainties in the \tmass\/ data, our results are
relatively insensitive to departures from universality.  For example,
the actual IR power law exponent among our sample stars could vary
significantly around the mean value of $-1.84$ used here, and we would
still find a very strong correlation between $k_{IR}$ and $R(V)$.  The
issue of the universality of IR extinction is best left to indepth
studies which utilize more focussed approaches and more appropriate
datasets, such as by Larson \& Whittet (2005) and Nishiyama et al.
(2006), both of which have found a range in values for the IR
exponent.

Values for $R(V)$ are often estimated from the formula $R(V) = 1.1
E(V-K)/E(B-V)$, which is based on van de Hulst's theoretical extinction
curve No. 15 (e.g., Johnson 1968).  In Figure \ref{figEVMINK} we plot
$E(V-K)/E(B-V)$ vs. $R(V)$ for our survey sightlines.  The dotted line
shows the van de Hulst relation, which agrees well with the data for
values of $R(V)$ near 3 --- not surprising since it was derived from a
theoretical curve with R = 3.05 --- but systematically deviates at
higher and lower values of $R(V)$.  The solid line in the figure shows
the best-fit linear relation which minimizes the residuals
perpendicular to the fit.  It is given by
\begin{eqnarray}
\label{eqnEKMINV}
R(V) = -0.26 + 1.19 \;  \frac{E(V-K)}{E(B-V)}  \;\;  .
\end{eqnarray}
This relationship was derived only from the sightlines with \tmass\/
$K$-band measurements (solid circles in Figure \ref{figEVMINK}) but
also agrees with those measurements based on Johnson $K$-band
photometry (open circles).  This exact form of this relation depends
slightly on our choice of an IR power law exponent of $-1.84$, and the
small scatter is another indicator that our data are consistent with a
single functional form for IR extinction.  The relationship in Equation
(\ref{eqnEKMINV}) can be reproduced by Equation (\ref{eqnIREXT}) with a
wavelength of $\lambda \simeq 2.1 \; \mu$m.

\subsubsection{$x_0$ vs. $\gamma$\label{secEXT_X0GAMMA}}

Among the many non-correlations between extinction quantities, one of the
most significant is that between the position of the peak of the 2175
\AA\/ bump (parametrized here by $x_0$) and its FWHM (parametrized by
$\gamma$).  The lack of a relationship between these two quantities, as
first reported in Paper I, places strong constraints on the nature of
the dust grains which produce the 2175 \AA\/ feature (see, e.g., Draine
2003).
    
Figure \ref{figGAMMAX0} shows a plot of $\gamma$ vs. $x_0$ for our
survey sightlines, along with their 1-$\sigma$ error bars.  As in all
previous studies, the lack of a correlation is clear.  

Figure \ref{figBUMPSTATS} shows the distribution of bump peak positions
(left panel) and widths (right panel) plotted in histogram form (shaded
regions).  With the exception of a few outliers, the distribution of
bump peaks can be fitted well with a Gaussian function, as indicated in
Figure \ref{figBUMPSTATS} by the smooth solid curve.  The centroid of
the Gaussian is at $x_0 =  4.5903$ \invmic\/ and its width is given by
$\sigma = 0.0191$ \invmic.  These correspond to a mean bump position of
2178.5 \AA\/ with a 1-$\sigma$ range of $\pm9.1$ \AA.  The RMS value of
the $x_0$ measurement errors for the full sample is $\pm0.0058$
\invmic\/ (corresponding to $\pm2.8$ \AA).  

While these results suggest small but significant variations in bump
positions --- as reported in Paper I --- the Gaussian-like distribution
of $x_0$ values in Figure \ref{figGAMMAX0} led us to consider that
perhaps our error analysis might underestimate the uncertainties in
$x_0$ and that the width of the Gaussian itself might represent the
true observational error.  We examined this issue by considering the
results for sightlines towards stars in open clusters and
associations.  We have previously used such sightlines to help estimate
extinction curve measurement uncertainties (Massa \& Fitzpatrick
1986).  If it is assumed that the true extinction curve is identical
for all cluster sightlines (based on the small spatial separation
between the sightlines), then each cluster curve represents an
independent measurement of the same curve and the variations from
sightline-to-sightline give the net measurement errors.  Since it is
unlikely that there is no cosmic variability among the sightlines, this
procedure provides upper limits on observational errors.  We utilized
data for the 13 clusters with five or more stars in our survey (NGC
457, Cr 463, NGC 869, NGC 884, NGC 1977, NGC 2244, NGC 3293, Tr 16, NGC
4755, NGC 6231, NGC 6530, Tr 37, and Cep OB3), yielding a total of 154
sightlines.  For each cluster we computed the mean value of $x_0$ and
subtracted it from the individual cluster values.  We then examined the
ensemble of residuals for the full cluster sample.  The shape of the
residuals distribution is Gaussian-like, although very slightly skewed
towards positive values.  A Gaussian fit yields the result shown by the
dotted curve in the $x_0$ panel of Figure \ref{figGAMMAX0} (scaled to
match the height of the main distribution), which has a width given by
$\sigma = 0.011$ \invmic.  This is larger than the RMS measurement
error for the cluster sample, i.e., $\pm0.0058$ \invmic\/, but
significantly smaller than the observed width of the full sample.  For
a number of the clusters, there are obvious curve-to-curve variations
present, and the width of the residuals distribution must certainly
overestimate the measurement errors.  From this more detailed analysis,
our conclusion remains that the small variations of $\pm9.1$ \AA\/ seen
in the full survey sample, are significantly larger than the expected
observational errors and indicate true variations in the position of
the bump peak.

The distribution of bump widths (right panel of Figure
\ref{figBUMPSTATS}) is decidedly non-Gaussian, with a strong tail in
the direction of large values.  The main peak of the distribution,
however, is Gaussian in appearance and a fit to this region (smooth
curve in the figure) yields a centroid of 0.890 \invmic\/ and a width
of $\sigma = 0.050$ \invmic.  The RMS value of the observational errors
is $\pm0.031$ \invmic, and so the width of this Gaussian, and the width
of the whole distribution is clearly larger than can be accounted for
by observational errors.  Again consistent with earlier results, we
find that the bump widths vary significantly from sightline to
sightline, but with no correlation with the centroid position.  The
shape of the $\gamma$ distribution might
suggestion two populations of bumps --- one characterized by the
Gaussian fit and the other characterized by a larger mean centroid
($\sim$1.1 \invmic) and a wider range of values ($\sigma \simeq 0.1$
\invmic), but this is not the only possible interpretation of the
results.  We examined the bump widths for the cluster sample as above.
However, the distribution of cluster residuals is complex, showing the
asymmetry of the full sample, and indicating significant variations
within the clusters, and we were thus not able to confirm the accuracy
of the measurement errors.  The range in observed $\gamma$ values is so
large, however, that the evidence for cosmic scatter is unambiguous.

\subsection{$R(V)$-Dependence\label{secEXT_RDEPEND}}
 
Cardelli, Clayton, \& Mathis (1988, 1989) were the first to demonstrate
a link between UV and optical/IR extinction by showing that $R(V)$ is
related to the level of UV extinction.  Essentially, sightlines with
large $R(V)$ values tend to have low UV extinction, and vice versa.
Cardelli et al. quantified this relationship in the following way:
\begin{eqnarray}
\label{eqnCCM1}
\frac{A(\lambda)}{A(V)} = a(\lambda) + b(\lambda) \; R(V)^{-1} \;\;\;\; ;
\end{eqnarray}
i.e., the total extinction at wavelength $\lambda$ normalized by the
total extinction at $V$ is a linear function of $R(V)^{-1}$.  In the
time since the original work, the perception of this relationship has
evolved to the point where it is often referred to as a ``law'' and
Galactic extinction curves are often stated or assumed to be a
1-parameter family (with $R(V)^{-1}$ as the parameter).  Recently, for
example, Valencic et al. (2004) found that 93\% of a large sample of
Galactic extinction curves obey a modified form of this relation.  In
this section, we will show that the relationship in Equation
(\ref{eqnCCM1}) is partially illusory and that Galactic extinction curves
are decidedly not a 1-parameter family in $R(V)^{-1}$.

The original basis for Equation (\ref{eqnCCM1}) is data such as shown
in Figure \ref{figRVAV}, where we plot $A(\lambda)/A(V)$ vs.
$R(V)^{-1}$ at four different UV wavelengths for our survey sightlines
(see Figure 1 in Cardelli et al. 1989).  It is clear why a linear
function would be chosen to quantify the obvious relationships seen in
the figure, and the data give the impression of being reasonably
well-correlated.  The solid line in the 2695 \AA\/ panel shows an
example of such a linear relationship. It is a weighted fit which
minimizes the residuals in the direction perpendicular to the fit, and
is given by $A(2695 \; {\rm \AA})/A(V) =  0.58 + 4.73 \; R(V)^{-1}$.
The appearance of Figure \ref{figRVAV} is, however, deceiving.  The
normalization used in the y-axis is constructed from the measured
values of $E(\lambda-V)/E(B-V)$ by the transformation \begin{eqnarray}
\label{eqnCCM2} \frac{A(\lambda)}{A(V)} \equiv
\frac{E(\lambda-V)}{E(B-V)} \; R(V)^{-1} + 1\;\;\;\; .  \end{eqnarray}
Thus, the four panels in Figure \ref{figRVAV} essentially amount to
plots of $xy$ vs. $x$ and, {\it even if $x$ and $y$ were completely
unrelated}, some degree of apparent correlation would inevitably
appear.  In addition, if there actually were an intrinsic relationship
between $x$ and $y$, its significance could be greatly
overinterpreted.
 
The true significance of the relationships plotted in Figure
\ref{figRVAV} could be fairly assessed if the measurement errors were
well-determined, including the effects introduced by the transformation
in Equation (\ref{eqnCCM2}).  Our Monte Carlo error analysis allows us
to quantify the uncertainties in any combination of fitted or measured
quantities and fully take into account the artificial correlation
induced by Equation (\ref{eqnCCM2}).  The principal axes of the
1-$\sigma$ errors for $A(\lambda)/A(V)$ and $R(V)^{-1}$ are indicated
in Figure \ref{figRVAV} by the dotted error bars and show the strong
correlation produced by the chosen normalization.  Ironically, because
of the normalization and the resultant error correlations, the
appearance of correlations in Figure \ref{figRVAV} is actually enhanced
by uncertainties in $R(V)$.  When the correlated errors are taken into
account, the linear relation shown in the 2695 \AA\/ panel of Figure
\ref{figRVAV} is found to have a reduced $\chi^2$ value of $\sim$4.2,
indicating that the scatter about the relation is more than four times
greater than accounted for by the known uncertainties in the data.

Although our error analysis offers us a way to overcome the
complications arising from the normalization chosen by Cardelli et al.,
a more direct approach to evaluating the relationship between UV and IR
extinction is simply to return to the actual quantities determined by
the extinction analysis and look at plots of $E(\lambda-V)/E(B-V)$ vs.
$R(V)$.  A small amount of algebra (i.e., combining Equations
[\ref{eqnCCM1}] and [\ref{eqnCCM2}]) will show that, if a linear
relationship as in Equation (\ref{eqnCCM1}) exists, then the following
relation must also hold:
\begin{eqnarray}
\label{eqnCCM3}
\frac{E(\lambda-V)}{E(B-V)} = b(\lambda) + [a(\lambda)-1] \; R(V) \;\;\;\; .
\end{eqnarray}
I.e., any true linear correlation between $A(\lambda)/A(V)$ and
$R(V)^{-1}$ will also be present between $E(\lambda-V)/E(B-V)$ and
$R(V)$, with a simple transformation relating the linear coefficients.
The great advantage to viewing the data in the $E(\lambda-V)/E(B-V)$
vs. $R(V)$ reference frame is that no artificial correlations (in either the parameters or their uncertainties) are
introduced and we only have to deal with the natural correlations which
arise from the dependence of all the extinction parameters on the
properties of the best-fit stellar SEDs.  Figure \ref{figRVKLAM}
illustrates this point.  It shows plots of $E(\lambda-V)/E(B-V)$ vs.
$R(V)$ (along with their associated 1-$\sigma$ errors) for the same
four wavelengths as in Figure \ref{figRVAV}.  The 2695 \AA\/ panel
shows a weighted linear fit, which is given by $E(2695 \; {\rm
\AA}-V)/E(B-V) = 4.80 - 0.44 \; R(V)$.  This is nearly exactly what
would be expected from the fit to $A(2695 \; {\rm \AA})/A(V)$ vs.
$R(V)^{-1}$ in Figure \ref{figRVAV} and the transformation in Equation
(\ref{eqnCCM3}).  The exact transformation is shown by the nearly
coincident dashed line in Figure \ref{figRVKLAM}, verifying the
argument leading to Equation (\ref{eqnCCM3}).  (A careful comparison
between the lines and the data points will show that the two panels do
indeed show the same relationship.)

From a statistical point of view, the presentations in Figures
\ref{figRVAV} and \ref{figRVKLAM} are identical, as are the linear fits
to the 2965 \AA\/ data.  In fact, both fits have nearly identical,
underwhelming, reduced $\chi^2$ values of $\sim$4.2.  Fits performed at
wavelengths shortward of 2675 \AA, show increasingly large values of
$\chi^2$ (e.g., $\chi^2 = 6.9$ for a linear fit at 1665 \AA).  The
lesson from Figures \ref{figRVAV} and \ref{figRVKLAM} is that the
choice of normalization affects the perception of how well-related
$R(V)$ is to the level of UV extinction.  The data in Figure
\ref{figRVAV} look better-correlated than do those in Figure
\ref{figRVKLAM} --- but they are not.   The ``cosmic scatter'' is
appreciable and there is no functional relationship between $R(V)$ and
the UV extinction (whether linear or more complex) for which the
scatter approaches the current level of measurement errors.  Although
there is a relationship in the sense that large-$R(V)$ curves [i.e., $R(V)
\gtrsim 4.0$] differ systematically from low-$R(V)$ curves, UV extinction
properties cannot be expressed as a 1-parameter family in $R(V)$ at
anywhere near the level of observational accuracy.

Note that the discussion above is not affected by the likelihood that
some of the sightlines are composites, possibly spanning distinct
regions of very different $R(V)$.  If a universal relationship of the
form in Equation (\ref{eqnCCM1}) were to hold, then composite regions
would still lie along the line $a(\lambda) + b(\lambda) \; R(V)^{-1}$
and would not result in ``cosmic scatter.''

The recent study by Valencic et al. (2004) examined the
$R(V)$-dependence of UV extinction by looking at correlations between
$R(V)^{-1}$ and UV fit parameters based on the Paper III formulation,
as had been done in F99 and F04.  As noted earlier, they concluded that
most Galactic curves (i.e., 93\%) are consistent with a
$R(V)$-dependent ``law.''  While the approach was somewhat different,
the results of the Valencic et al. study suffer the same problem as
shown in Figures \ref{figRVAV} and \ref{figRVKLAM} because they
explicitly multiply the UV fit coefficients by $R(V)^{-1}$, thus
forcing a correlation in the errors and enhancing the perception of a
correlation between the quantities.  The large percentage of curves
believed to be consistent with a single $R(V)^{-1}$ relation results
from the complications introduced by the choice of curve
normalization.  Figure \ref{figRVFM} shows the relationship between
$R(V)$ and the UV fitting parameters in their original form, in which
the correlations among the parameters and their errors are not
artificially enhanced.  The figure contains several extra quantities
derived from the fit parameters which help describe features of the UV
curves and which are defined in \S \ref{secMETHOD} (and also in the
figure legend).  This figure again shows that, while systematic
differences exist between high-$R(V)$ and low-$R(V)$ curves, there is
no simple (or complex) relation between $R(V)$ and the UV fitting
properties which is consistent with measurement errors.


\section{SUMMARY\label{secDISCUSS}}

Several of our findings are worth emphasizing.  

\noindent {\underline {\bf Variability of \boldmath{$x_0$} and
\boldmath{$\gamma$}:}}$\;$ We found, in accordance with our previous
analysis, that while the central position of the 2175 \AA\/ bump does not
vary much, it is indeed variable.  On the other hand, the bump width
varies considerably.  Further, there still appears to be no
relationship between the central position of the bump and and its width
-- verifying the results described in Paper I to a higher degree of
accuracy than previously possible.
  
\noindent {\underline {\bf Correlations among the fit parameters:}}$\;$
Generally, the various fit parameters which describe the shape of the
UV-through-IR extinction curves are not related to one another.  The
only exceptions are $c_1-c_2$ and $k_{IR}-R(V)$.  The first relates the
slope and intercept of the UV portion of the curve and is effectively a
functional relationship for most of the data.  The latter relates the
scale factor of the power law used to describe the IR portion of the
curve and the ratio of total-to-selective extinction.  At the level of
our measurements errors, these two parameters are functionally related and
consistent with a universal form to IR extinction.  However, due to observational uncertainties, the dominance of diffuse ISM sightlines in our study, and the limitations of the IR data we employ, our results
are not ideal for addressing the detailed shape of IR extinction and
significant sightline-to-sightline variations could still exist.

\noindent {\underline {\bf Correlations with dust density:}}$\;$ As
determined in the past, various properties of the extinction curves are
weakly correlated with the mean line of sight dust density, as measured
by $E(B-V)/d$.  Presumably, these correlations reflect the operation of
a physical process, such as grain growth or coagulation.
  
\noindent {\underline {\bf Correlations of \boldmath{$R(V)$} with UV
curve properties relationship:}$\;$ Correlations between $R(V)$ and UV
curve properties have received considerable attention over the years
since Cardelli et al. (1988) first pointed out that a large-scale trend
could be found.  By expressing the curves in their native form, we
verify that a there is a weak relationship between $R(V)$ and the UV in
the sense that sightlines with extremely large $R(V)$ values tend to
have low normalized UV extinction curves.  However, this relationship
is only evident for the largest $R(V)$ sightlines.  For the majority of
sightlines, there is no evidence for an $R(V)$-dependence of the
extinction curve shapes.  Specifically, inspection of Figure
\ref{figRVKLAM} shows that for diffuse ISM sightlines ($2.4 < R(V) <
3.6$), which comprise the bulk (82\%) of our sample, no relationship
exists, even though a large range in the extinction is present.
Moreover, Figure \ref{figRVFM} shows that the 2175 \AA\/ bump
properties display a similar behavior, i.e., a trend is evident only
for the largest $R(V)$ values, with the strength of the bump (as
measured by $A_{bump}$ or $E_{bump}$) tending to be slightly weaker
than the average for lower $R(V)$ sightlines.  Thus, we conclude that
there is no global 1-parameter family of extinction curves, although
extremely large $R(V)$ curves tend to have distinctive properties.
 
\noindent {\underline {\bf Extinction curve variability -- the meaning
and utility of an average curve:}}$\;$ The previous result begs the
meaning of an average extinction curve.  As noted in
\S \ref{secEXT_PROPERTIES}, simple mean curves always reflect the biases
of their parent samples.  Our mean curve for the diffuse ISM (Figure
\ref{figAVGCURVE}, an average of all curves with $2.4 \leq R(V) \leq
3.6$, and that from Valencic et al.\ (2004, for R=3.1) are derived from
the largest samples and probably provide the best estimate of mean
Galactic diffuse ISM extinction properties at short wavelengths.
However, one must always be mindful of the dark shaded region in Figure
\ref{figAVGCURVE} which illustrates the RMS variance that can be
expected for an extinction curve along an arbitrary diffuse sightline.
Typical RMS dispersions in $k(\lambda-V)$ are 0.31, 0.68, 0.62, 1.44
at $\lambda = $2695, 2175, 1665, 1245 \AA, respectively.  This means
that if the mean curve is used to deredden an object with $E(B-V) =
0.50$~mag, the uncertainty in the dereddened continuum at these
wavelengths would be 0.15, 0.30, 0.30 and 0.72~mag, respectively, due
to uncertainties in the extinction alone!  It is, however, possible to
take advantage of localized uniformity in the extinction to reduce this
error (e.g., Massa \& Savage 1981).

\noindent {\underline {\bf Physical implications:}}$\;$ Perhaps the
best way to summarize the physical origin of the $R(V)$-dependence for
Galactic extinction is to examine the $R(V)^{-1} - c_2$ plot shown in
Figure~\ref{figRVC2}, which includes a best-fit linear relation (such a
relation was the basis for the $R(V)$-dependent curves produced by F99
and F04 -- see Figure~10 of F04).  Figure~\ref{figRVC2} can be
summarized thusly: when extinction curves are steep in the optical
(large $R(V)^{-1}$) they tend to stay steep in the UV (large $c_2$) and
when extinction curves are flat in the optical (small $R(V)^{-1}$) they
tend to stay flat in the UV (small $c_2$).  However, as the scatter in
the Figure illustrates, this trend is only apparent for extreme values
of $R(V)^{-1}$ or $c_2$.  In general, there is no unique relation
between these parameters over the range spanned by most of the sample.
It is likely that the general connection between UV and optical
extinction slopes simply reflects the fact that the overall grain size
distribution affects all wavelengths.  But the presence of such a large
scatter demonstrates that several other factors (e.g., chemical
composition, grain history, coagulation, coating, radiation
environment, etc.) must also be involved.  In other words, the large
variance in the relation between UV and optical slopes indicates that
dust grain size distributions do not behave as a 1-parameter family.

\noindent {\underline {\bf Final remarks:}}$\;$  Having painted a
negative picture of the relationship between $R(V)$ and extinction
properties at other wavelengths, it would be disingenuous of us to
present yet another set of $R(V)$-dependent curves.  The results of
Cardelli et al. (1989), F04 (which supersedes those of F99), and
Valencic et al. (2004) are all reasonable, and the differences among
them are instructive of the biases introduced by sample selection and
methodology.  It should always be remembered that these curves
represent very general trends in Galactic extinction and do not
constitute a standard of normalcy.  Finally, although the prospect of
accurately dereddening an unknown SED using a mean extinction curve is
poor, this same variability demonstrates that extinction curves are
responsive to local conditions, so that each one contains potentially
unique information about the grains along the sightline.


\begin{acknowledgments}
E.F. acknowledges support from NASA grant NAG5-12137, NAG5-10385, and
NNG04GD46G.  D.M. acknowledges support from NASA grant NNG04EC01P. We
are grateful to the referee Geoff Clayton for helpful comments and
suggestions. Some of the data presented in this paper were obtained
from the Multimission Archive at the Space Telescope Science Institute
(MAST).  STScI is operated by the Association of Universities for
Research in Astronomy, Inc., under NASA contract NAS5-26555. Support
for MAST for non-HST data is provided by the NASA Office of Space
Science via grant NAG5-7584 and by other grants and contracts.  This
publication also makes use of data products from the Two Micron All Sky
Survey, which is a joint project of the University of Massachusetts and
the Infrared Processing and Analysis Center/California Institute of
Technology, funded by the National Aeronautics and Space Administration
and the National Science Foundation.
  \end{acknowledgments}


\appendix
\section{Obtaining the Data from this Study}

The extinction curves and associated information from this study are
available via anonymous ftp at ``ftp.astronomy.villanova.edu''.  After
logging in as ``anonymous'', change to the appropriate directory by
typing ``cd fitz$/$FMV\_EXTINCTION''.  A ``README'' file, two IDL
procedure files, and two tar files are present.  The tar files contain
the extinction curves from this program  and the R-dependent curves
from Fitzpatrick 2004 (``FMV\_EXTCURVES.tar'' and ``F04\_RCURVES.tar'',
respectively).  The tar files unfold into individual ascii data files
for each extinction curve, containing the 12 extinction parameters
which describe the curve and the curve itself (along with 1-$\sigma$
uncertainties for the curves from this paper).  The two IDL procedures
provide tools for reading data from the ascii files and for
constructing extinction curves from the 12 fit parameters
(``READ\_FMV\_FILES.pro'' and ``MAKE\_FMV\_CURVE.pro'', respectively).


\clearpage
\bibliographystyle{apj}

\clearpage



\begin{deluxetable}{lcrcrrc} 
\tabletypesize{\scriptsize}
\tablenum{1} 
\tablewidth{0pc} 
\tablecaption{Basic Data for Survey Stars [The complete version of this
table is in the electronic edition of the Journal.  The printed edition
contains only a sample.]
\label{tabSTARS}}
\tablehead{ 
\colhead{Star\tablenotemark{a}}         &
\colhead{Spectral}                      & 
\colhead{V}                             &
\colhead{Distance\tablenotemark{c}}     &
\colhead{$l$}                &
\colhead{$b$}                &
\colhead{Reference}                     \\
\colhead{}                         &  
\colhead{Type\tablenotemark{b}}    & 
\colhead{(mag)}                    &
\colhead{(pc)}                     &
\colhead{($\degr$)}                & 
\colhead{($\degr$)}                & 
\colhead{}                         }
\startdata
HD698 &  B5 II: SB & 7.10 & 1125 & $117.689$ & $ -4.25$ &  1  \\
HD3191 &  B1 IV:nn & 8.58 & 1203 & $121.068$ & $ -1.36$ &  2  \\
BD+57 245 (NGC 457) &  \nodata & 9.85 & 2429 & $126.583$ & $ -4.58$ & \nodata \\
BD+57 252 (NGC 457) &  B1 IV & 9.51 & 2429 & $126.644$ & $ -4.42$ &  2  \\
NGC 457 Pesch 34 &  \nodata & 10.61 & 2429 & $126.646$ & $ -4.38$ & \nodata \\
NGC 457 Pesch 13 &  \nodata & 10.78 & 2429 & $126.646$ & $ -4.38$ & \nodata \\
NGC 457 Pesch 9 &  B1 V & 9.83 & 2429 & $126.646$ & $ -4.38$ &  3  \\
Cr 463 \#18 &  \nodata & 10.35 & \phn702 & $127.091$ & $  9.20$ & \nodata \\
BD+70 131 (Cr 463) &  \nodata & 10.06 & \phn702 & $127.280$ & $  9.17$ & \nodata \\
Cr 463 \#5 &  \nodata & 10.37 & \phn702 & $127.264$ & $  9.37$ & \nodata \\
\enddata
\tablenotetext{a}{The stars are listed in order of increasing Right Ascension using the most commonly adopted forms of their names. The first preference was ``HDnnn'', followed by ``BDnnn'', etc.  There are 185 survey stars which are members of open clusters or associations, or 56\% of the sample.  The identity of the cluster or association is either contained in the star name itself (e.g., NGC 457 Pesch 34) or is given in parentheses after the star's name.}
\tablenotetext{b}{Spectral types were selected from those given in the SIMBAD database, and the source of the adopted types is shown in the ``Reference'' column.  When multiple types were available for a particular star, we selected one based on our own preferred ranking of the sources.}
\tablenotetext{c}{The NGC 2244 distance is from Perez et al. 1987; the NGC 3293 distance is from Bolona \& Crampton 1974; the Trumpler 14 and 16 distances are from Massey \& Johnson 1993; the Cep OB3 distance is from  Crawford \& Barnes 1970.  The distances to all other clusters or associations are from the Open Clusters and Galactic Structure database maintained by Wilton S. Dias, Jacques L\'{e}pine, Bruno S. Alessi, and Andr\'{e} Moitinho at http://www.astro.iag.usp.br/~wilton/.  For the non-cluster stars, distances were calculated using the $E(B-V)$ values from this study and the absolute magnitudes from Turner 1980 (for mid-B and earlier types) and Blaauw 1963 (for mid-B and later types).}
\tablerefs{(1) Hiltner 1956; (2) Morgan, Code, \& Whitford 1955; (3) Hoag, \& Applequist 1965; (4) Ma\'{i}z-Apell\'{a}niz, Walborn, Galu\'{e}, \& Wei 2005; (5) Johnson, \& Morgan 1955; (6) Slettebak 1968; (7) Schild 1965; (8) Racine 1968; (9) Morgan, Keenan, \& Kellman 1943; (10) Mendoza 1956; (11) Morgan, Whitford, \& Code 1953; (12) Osawa 1959; (13) Cowley, Cowley, Jaschek, \& Jaschek 1969; (14) Metreveli 1968; (15) Guetter 1968; (16) Boulon, \& Fehrenbach 1959; (17) Buscombe 1962; (18) Roman 1955; (19) Georgelin, Georgelin, \& Roux 1973; (20) Bouigue 1959; (21) Wenzel 1951; (22) Smith 1972; (23) Penston, Hunter, \& O'Neill 1975; (24) McNamara 1976; (25) Levato H., \& Abt H.A.; (26) Schild, \& Chaffee 1971; (27) Borgman 1960; (28) Warren, \& Hesser 1977; (29) Sharpless 1952; (30) Racine 1968; (31) Crawford, Limber, Mendoza, Schulte, Steinman, \& Swihart 1955; (32) Johnson, \& Morgan 1953; (33) Meadows 1961; (34) Johnson 1962; (35) Hoag, \& Smith 1959; (36) Barbier, \& Boulon 1960; (37) Young 1978; (38) Claria 1974; (39) Feast, Thackeray, \& Wesselink 1955; (40) Moffat, \& Fitzgerald 1974; (41) Houk 1978; (42) Feast, Stoy, Thackeray,\& Wesselink 1961; (43) Denoyelle 1977; (44) Hoffleit 1956; (45) Turner, Grieve, Herbst, \& Harris 1980; (46) Morrell, Garcia, \& Levato 1988; (47) Massey, \& Johnson 1993; (48) Levato, Malaroda, Morrell, Garcia, \& Hernandez 1991; (49) Morris 1961; (50) Hiltner, Garrison, \& Schild 1969; (51) Houk, \& Cowley 1975; (52) Seidensticker 1989; (53) Feast, Thackeray, \& Wesselink 1957; (54) Perry, Franklin, Landolt, \& Crawford 1976; (55) Schild 1970; (56) De Vaucouleurs 1957; (57) Walraven, \& Walraven 1960; (58) Garrison 1967; (59) Buscombe 1969; (60) Hardie, \& Crawford 1961; (61) Perry, Hill, \& Christodoulou 1991; (62) Schild, Neugebauer, \& Westphal 1971; (63) Schild, Hiltner, \& Sanduleak 1969; (64) Levato, \& Malaroda 1980; (65) Garrison, \& Schild 1979; (66) Roman 1956; (67) Houk, \& Smith-Moore 1988; (68) Hiltner, Morgan, \& Neff 1965; (69) Walker 1957; (70) Walker 1961; (71) Stebbins, \& Kron 1956; (72) Guetter 1964; (73) Hill 1970; (74) Herbig, \& Spalding 1955; (75) Vrba, \& Rydgren 1984; (76) Hack 1953; (77) Roman 1951; (78) Divan 1954; (79) Garrison, \& Kormendy 1976; (80) Simonson 1968; (81) Blaauw, Hiltner, \& Johnson 1959; (82) Pecker 1953; (83) Boulon, Duflot, \& Fehrenbach 1958}
\end{deluxetable}
\clearpage


\begin{deluxetable}{lc} 
\tablewidth{0pc} 
\tablecaption{Adopted Temperature Scale for Main Sequence O Stars
\label{tabOTEMPS}}
\tablenum{2} 
\tablehead{ 
\colhead{Spectral}      &
\colhead{$T_{eff}$}     \\
\colhead{Type}          &  
\colhead{(K)}            }
\startdata
O6    &  $40000$  \\
O6.5  &  $38500$  \\
O7    &  $37000$  \\
O7.5  &  $36500$  \\
O8    &  $36000$  \\
O8.5  &  $34750$  \\
O9    &  $33500$  \\
O9.5  &  $32750$  \\
B0    &  $32000$  \\
\enddata
\end{deluxetable}
\clearpage


\begin{deluxetable}{llcccrc} 
\tabletypesize{\footnotesize}
\tablenum{3} 
\tablewidth{0pc} 
\tablecaption{Best-Fit Parameters for Survey Stars [The complete version of this
table is in the electronic edition of the Journal.  The printed edition
contains only a sample.]
\label{tabPARMS}}
\tablehead{ 
\colhead{Star}              &
\colhead{$T_{eff}$\tablenotemark{a}}         & 
\colhead{$\log g$\tablenotemark{b}}          &
\colhead{[m/H]\tablenotemark{c}}             & 
\colhead{$v_{turb}$\tablenotemark{d}}        & 
\colhead{$\theta_R$}        &
\colhead{$E(B-V)$}          \\
\colhead{}                  &  
\colhead{(K)}               & 
\colhead{}                  & 
\colhead{}                  & 
\colhead{$\rm (km/s)$}      & 
\colhead{(mas)}             &
\colhead{(mag)}             }
\startdata
HD698 & $ 18434\pm 499$ & $ 3.72\pm 0.31$ & $-0.18\pm 0.10$ & $  3.2\pm  0.5$ & $0.0670\pm0.0014$ & $ 0.37\pm 0.01$ \\
HD3191 & $ 24001\pm 890$ & $ 3.41\pm 0.28$ & $-0.43\pm 0.06$ & $10$ & $0.0328\pm0.0009$ & $ 0.68\pm 0.01$ \\
BD+57 245 (NGC 457) & $ 22885\pm 697$ & $ 3.64\pm 0.20$ & $-0.39\pm 0.08$ & $  5.5\pm  0.6$ & $0.0160\pm0.0004$ & $ 0.50\pm 0.01$ \\
BD+57 252 (NGC 457) & $ 24924\pm 616$ & $ 3.64\pm 0.16$ & $-0.47\pm 0.06$ & $10$ & $0.0173\pm0.0004$ & $ 0.51\pm 0.01$ \\
NGC 457 Pesch 34 & $ 23594\pm 661$ & $ 3.91\pm 0.17$ & $-0.07\pm 0.06$ & $  1.2\pm  0.8$ & $0.0110\pm0.0003$ & $ 0.51\pm 0.01$ \\
NGC 457 Pesch 13 & $ 22023\pm 809$ & $ 3.84\pm 0.17$ & $-0.47\pm 0.11$ & $  2.1\pm  0.8$ & $0.0114\pm0.0003$ & $ 0.51\pm 0.01$ \\
NGC 457 Pesch 9 & $ 25738\pm 577$ & $ 3.77\pm 0.16$ & $-0.75\pm 0.07$ & $  9.1\pm  0.7$ & $0.0147\pm0.0003$ & $ 0.54\pm 0.01$ \\
Cr 463 \#18 & $ 11891\pm 286$ & $ 3.99\pm 0.16$ & $-0.60\pm 0.10$ & $0$ & $0.0198\pm0.0003$ & $ 0.35\pm 0.01$ \\
BD+70 131 (Cr 463) & $ 11351\pm 204$ & $ 3.93\pm 0.17$ & $-0.73\pm 0.07$ & $  5.1\pm  0.4$ & $0.0210\pm0.0003$ & $ 0.28\pm 0.01$ \\
Cr 463 \#5 & $ 11859\pm 230$ & $ 4.10\pm 0.19$ & $-0.59\pm 0.09$ & $  2.0\pm  0.4$ & $0.0170\pm0.0002$ & $ 0.30\pm 0.01$ \\
\enddata
\tablenotetext{a}{For the O stars analyzed using the TLUSTY atmosphere models, the values of $T_{eff}$ were adopted from the spectral type vs. $T_{eff}$ relation given in Table 2.  These stars can be identified by their 1-$\sigma$ uncertainties, which are $\pm$1000 K.}

\tablenotetext{b}{For stars in clusters, the surface gravities are determined as discussed in \S 3.1 and rely on stellar evolution models and cluster distance determinations.  Surface gravities for non-cluster stars are not always well-determined, because of a lack of specific spectroscopic indicators.  In some cases, the best-fit solutions for these stars indicated physically unlikely results (i.e.,  $\log g >$  $\sim$4.3 or $\log g <$ $\sim$3.0).  For these stars, a value of $\log g$ = 3.9 was assumed (which is the mean $\log g$ of the rest of the sample) and a 1-$\sigma$ uncertainty of $\pm$0.2 was incorporated in the error analysis.  These cases can be identified by $\log g$ entries of ``$3.9\pm0.2$.''}

\tablenotetext{c}{For the O stars in the sample, our fitting procedure utilized solar abundance TLUSTY models.  For these stars the values of [m/H] are indicated by entries of ``0'' without uncertainties.}

\tablenotetext{d}{For the O stars, the adopted TLUSTY models incorporate $v_{turb} = 10$ km\,s$^{-1}$.  For these stars the values of $v_{turb}$ are indicated by entries of ``10'' without uncertainties.  For the B stars, which were modeled using ATLAS9 models, the values of $v_{turb}$ were determined by the fitting procedure, but were constrained to lie between 0 and 10 km\,s$^{-1}$. Stars whose best-fit SED models required these limiting values are indicated by $v_{turb}$ entries of ``0'' or ``10'', without error bars. The uncertainties for stars with best-fit $v_{turb}$ values close to these limits may be underestimated due to this truncation.}
\end{deluxetable}
\clearpage


\topmargin 0.6in
\begin{deluxetable}{lccrcrccccrrccc}
\rotate 
\tabletypesize{\scriptsize}
\tablenum{4} 
\tablewidth{0pc} 
\tablecaption{Best-Fit Extinction Curve Parameters for Survey Stars
[The complete version of this table is in the electronic edition of the
Journal.  The printed edition contains only a sample.] 
\label{tabEXTINCT}}
\tablehead{ 
\colhead{}                  &
\multicolumn{7}{c}{UV Coefficients\tablenotemark{a}}      &
\colhead{}                  &
\multicolumn{3}{c}{Optical Spline Points\tablenotemark{b}}   &
\colhead{}                  &
\multicolumn{2}{c}{IR Coefficients\tablenotemark{c}} \\ \cline{2-8} \cline{10-12} \cline{14-15}  
\colhead{Star}              &
\colhead{$x_0$}             & 
\colhead{$\gamma$}          & 
\colhead{$c_1$}             & 
\colhead{$c_2$}             & 
\colhead{$c_3$}             & 
\colhead{$c_4$}             & 
\colhead{$c_5$}             &  
\colhead{}                  &
\colhead{$O_1$}             & 
\colhead{$O_2$}             & 
\colhead{$O_3$}             & 
\colhead{}                  &
\colhead{$R(V)$}            &
\colhead{$k_{IR}$}          } 
\startdata
HD698 & $ 4.551\pm 0.006$ & $ 0.96\pm 0.03$ & $ 0.07\pm 0.19$ & $ 0.99\pm 0.05$ & $ 2.95\pm 0.19$ & $ 0.15\pm 0.05$ & $ 6.51\pm 0.29$ &  & $ 2.38\pm 0.09$ & $ 1.33$ & $ 0.00$ &  & $ 3.94\pm 0.16$ & $ 1.70\pm 0.18$ \\
HD3191 & $ 4.636\pm 0.003$ & $ 0.94\pm 0.02$ & $-0.79\pm 0.20$ & $ 1.00\pm 0.04$ & $ 2.99\pm 0.16$ & $ 0.31\pm 0.02$ & $ 5.76\pm 0.09$ &  & $ 2.04\pm 0.06$ & $ 1.33$ & $ 0.01$ &  & $ 2.81\pm 0.09$ & $ 0.93\pm 0.14$ \\
BD+57 245 (NGC 457) & $ 4.561\pm 0.005$ & $ 0.89\pm 0.03$ & $-0.47\pm 0.31$ & $ 0.88\pm 0.06$ & $ 3.13\pm 0.19$ & $ 0.24\pm 0.03$ & $ 5.71\pm 0.16$ &  & $ 2.05\pm 0.06$ & $ 1.31$ & $ 0.00$ &  & $ 2.97\pm 0.13$ & $ 0.96\pm 0.17$ \\
BD+57 252 (NGC 457) & $ 4.577\pm 0.005$ & $ 0.92\pm 0.02$ & $-0.66\pm 0.25$ & $ 0.95\pm 0.05$ & $ 3.39\pm 0.17$ & $ 0.17\pm 0.02$ & $ 5.48\pm 0.23$ &  & $ 2.15\pm 0.06$ & $ 1.32$ & $ 0.00$ &  & $ 2.97\pm 0.10$ & $ 0.98\pm 0.15$ \\
NGC 457 Pesch 34 & $ 4.579\pm 0.003$ & $ 0.91\pm 0.01$ & $-0.55\pm 0.20$ & $ 0.86\pm 0.04$ & $ 3.32\pm 0.11$ & $ 0.31\pm 0.02$ & $ 6.14\pm 0.09$ &  & $ 1.99\pm 0.07$ & $ 1.30$ & $ 0.01$ &  & $ 2.94\pm 0.11$ & $ 0.84\pm 0.16$ \\
NGC 457 Pesch 13 & $ 4.587\pm 0.004$ & $ 0.87\pm 0.02$ & $-0.69\pm 0.33$ & $ 0.84\pm 0.07$ & $ 2.91\pm 0.14$ & $ 0.26\pm 0.02$ & $ 5.59\pm 0.17$ &  & $ 2.03\pm 0.07$ & $ 1.31$ & $ 0.01$ &  & $ 3.11\pm 0.11$ & $ 0.90\pm 0.16$ \\
NGC 457 Pesch 9 & $ 4.578\pm 0.006$ & $ 1.07\pm 0.03$ & $-1.13\pm 0.22$ & $ 1.08\pm 0.04$ & $ 4.31\pm 0.22$ & $ 0.18\pm 0.02$ & $ 5.00\pm 0.18$ &  & $ 2.37\pm 0.05$ & $ 1.35$ & $ 0.01$ &  & $ 2.76\pm 0.07$ & $ 0.79\pm 0.12$ \\
Cr 463 \#18 & $ 4.606\pm 0.009$ & $ 1.18\pm 0.05$ & $-0.47\pm 0.26$ & $ 1.03\pm 0.07$ & $ 5.23\pm 0.45$ & $ 0.30\pm 0.03$ & $ 5.09\pm 0.20$ &  & \nodata & $ 1.37$ & $ 0.01$ &  & $ 3.38\pm 0.17$ & $ 1.29$ \\
BD+70 131 (Cr 463) & $ 4.577\pm 0.011$ & $ 1.11\pm 0.05$ & $ 0.17\pm 0.32$ & $ 0.86\pm 0.08$ & $ 5.11\pm 0.49$ & $ 0.35\pm 0.04$ & $ 5.00\pm 0.19$ &  & \nodata & $ 1.37$ & $ 0.00$ &  & $ 3.40\pm 0.20$ & $ 1.30$ \\
Cr 463 \#5 & $ 4.610\pm 0.008$ & $ 1.08\pm 0.04$ & $ 0.45\pm 0.33$ & $ 0.74\pm 0.08$ & $ 4.77\pm 0.34$ & $ 0.32\pm 0.03$ & $ 5.06\pm 0.18$ &  & \nodata & $ 1.37$ & $ 0.00$ &  & $ 2.91\pm 0.18$ & $ 0.99$ \\
\enddata
\tablenotetext{a}{For the stars HD237019, HD18352, and HD25443 the long wavelength {\it IUE} spectra are incomplete.  For these cases we constrained the UV linear extinction component to follow the relation $c_1 = 2.18 - 2.91 c_2$ from Fitzpatrick 2004.  For these stars we list uncertainties for the $c_2$ values but not for the $c_1$ values.}

\tablenotetext{b}{The uncertainties in the $O_2$ and $O_3$ optical spline points (at wavelengths of 4000 \AA\/ and 5530 \AA, respectively) are typically 0.01 or less and are not listed.  For several stars --- those without $U$ band photometry --- we did not solve for the $O_1$ point at 3300 \AA.}

\tablenotetext{c}{For field stars without IR photometry, we assumed $R(V) = 3.1$ and $k_{IR} = 1.11$, with the latter based on the relation $k_{IR} = 0.63R(V) - 0.84$ from Fitzpatrick 2004.  For such stars in clusters, we adopted the mean $R(V)$ of the other cluster members and a value of $k_{IR}$ based on the aforementioned relation.  These assumed values are listed in the Table without uncertainties.  Several survey stars have apparently noisy {\it JHK} data and yielded very uncertain values of $k_{IR}$.  For these, we ultimately derived the extinction curve by solving for the best-fit value of $R(V)$ with $k_{IR}$ constrained to follow the Fitzpatrick 2004 relation.  The resultant $R(V)$ values are listed with their uncertainties while the $k_{IR}$ values are listed without uncertainties.}
\end{deluxetable}
\clearpage


\topmargin 0.0in
\begin{deluxetable}{cl}
\tablenum{5} 
\tablewidth{0pc} 
\tablecaption{Extinction Curve Parameters for Mean Curve in Figure 9
\label{tabAVGCURVE}}
\tablehead{ 
\colhead{Parameter}       &
\colhead{Value}          } 
\startdata
$x_0$    &  \phm{$-$}4.592 \invmic  \\
$\gamma$ &  \phm{$-$}0.922 \invmic  \\
$c_1$    &         $-0.175$         \\
$c_2$    &  \phm{$-$}0.807          \\
$c_3$    &  \phm{$-$}2.991          \\
$c_4$    &  \phm{$-$}0.319          \\
$c_5$    &  \phm{$-$}6.097          \\
$O_1$    &  \phm{$-$}2.055          \\
$O_2$    &  \phm{$-$}1.322          \\
$O_3$    &  \phm{$-$}0.000          \\
$R(V)$   &  \phm{$-$}3.001          \\
$k_{IR}$ &  \phm{$-$}1.057          \\
\enddata
\end{deluxetable}
\clearpage


\clearpage

\begin{figure}[ht]
\epsscale{1.0}
\plotone{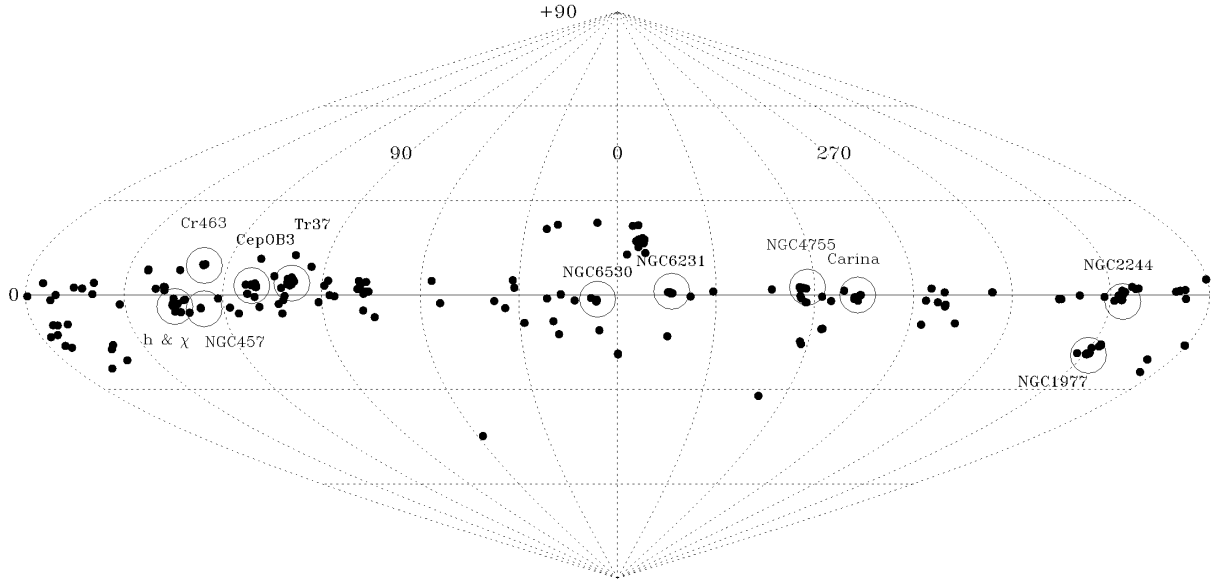}
\caption{Sky distribution of survey stars, shown in a sinusoidal
projection of Galactic coordinates.  The Galactic plane is the solid
horizontal line and the Galactic center is in the middle of the
figure.  The locations of open clusters or associations containing five
or more survey stars are indicated by the large circles, labeled with
the cluster name.  The sizes of the circles do not represent the
physical extent of the clusters or associations.
\label{figCOORDS}}
\end{figure}
\clearpage

\begin{figure}[ht]
\epsscale{0.50}
\plotone{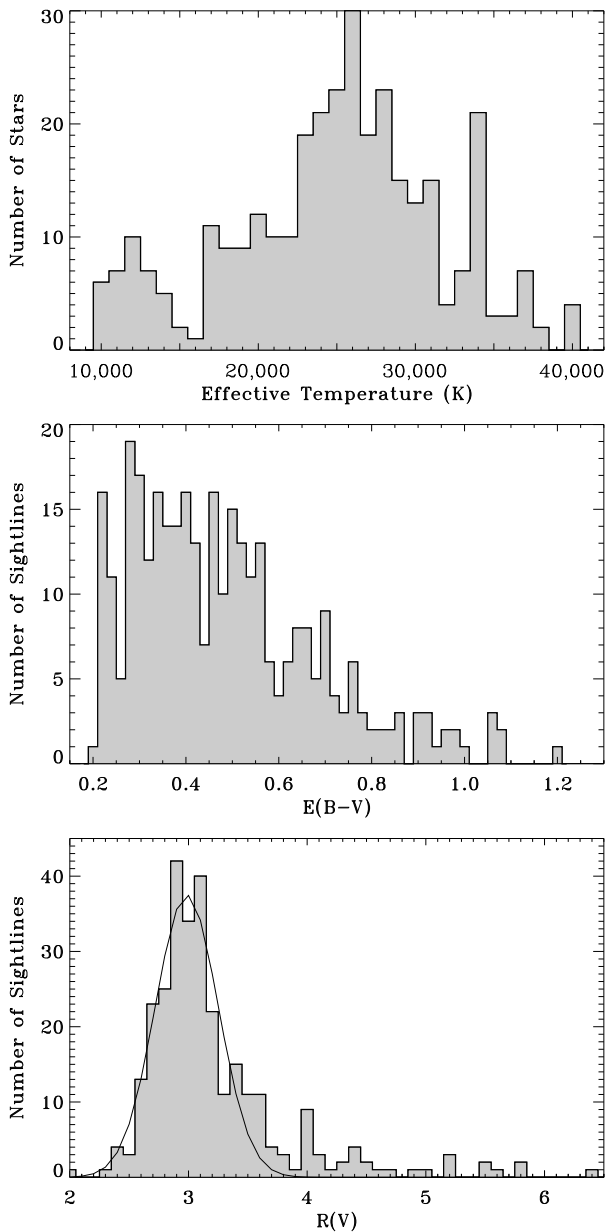}
\caption{Representative properties of the survey stars.  The three
panels show results from the analysis presented in \S \ref{secATLAS},
which help characterize the properties of our sample.  The typical
survey member is a mid-B star (top panel), with a median reddening of
$\overline{\ebv} = 0.45$ mag (middle panel), viewed along a
line-of-sight passing through the diffuse ISM (bottom panel).  The
bottom panel also shows a Gaussian fit to the values of $R(V) \equiv
A_V/\ebv$ in the neighborhood of the peak in the distribution.  The
peak is located at $R(V) = 2.99$, similar to the mean values usually
attributed to the diffuse ISM, and the width of the Gaussian
corresponds to $\sigma = \pm 0.27$.  The mean and median values of
$R(V)$ for the whole sample are 3.22 and 3.05, respectively.
\label{figSTARSTATS}}
\end{figure}
\clearpage

\begin{figure}[ht]
\epsscale{1.0}
\plotone{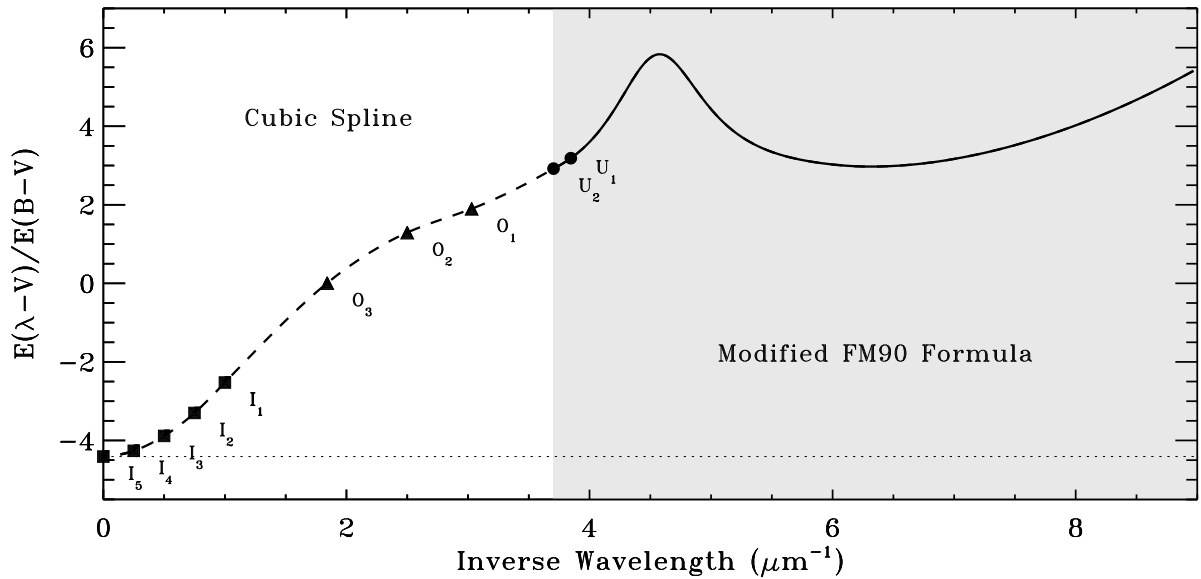}
\caption{A parametrized representation of normalized UV-through-IR
extinction (solid and dashed curve).  The curve consists of two parts:
1) $\lambda \leq 2700$ \AA\/ (shaded region), where we adopt a modified
version of the 3-component parametrization scheme of Paper III; and 2)
$\lambda > 2700$ \AA,  where we adopt a cubic spline interpolation
through sets of IR ($I_n$), optical ($O_n$), and UV ($U_n$) ``anchor
points.''  The values of the anchor points and the seven parameters
describing the UV portion of the curve are determined by fitting the
observed SED of a reddened star, as described in \S \ref{secMETHOD}.
The particular curve shown in this figure corresponds to that derived
for the star HD147933.
\label{figEXTCRV}}
\end{figure}
\clearpage

\begin{figure}[ht]
\epsscale{0.70}
\plotone{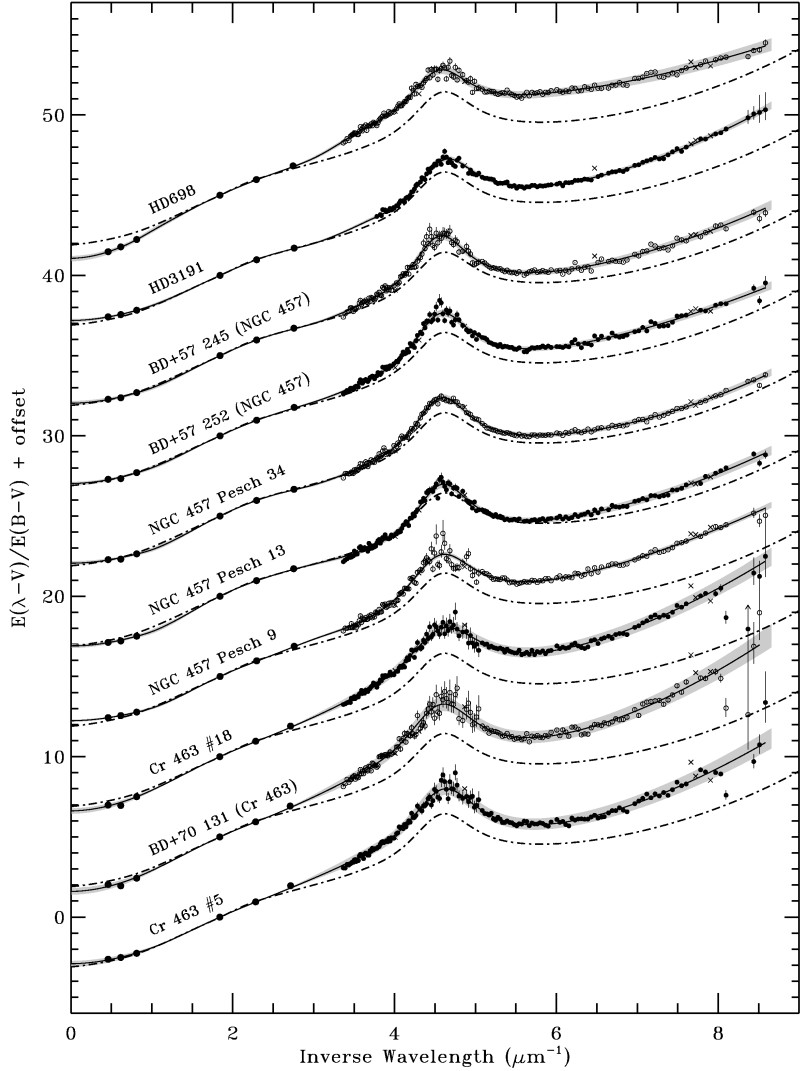}
\caption{Normalized extinction curves for 328 Galactic stars derived
using the extinction-without-standards approach.  The symbols show the
normalized ratios of the model atmosphere fluxes to 1) the \iue\/
spectrophotometry in the UV ($\lambda^{-1} > 3.3 \; \invmic$), 2) Johnson
{\it UBV} photometry in the optical, and 3) Johnson or \tmass\/ {\it
JHK} photometry in the near-IR ($\lambda^{-1} <  1 \; \invmic$).
Individual 1-$\sigma$ error bars are shown for the data points, but are
typically only visible in the region of the 2175 \AA\/ bump and in the
far-UV, where the signal level of the \iue\/ data is lowest. Small
crosses indicate \iue\/ data points excluded from the fit, for the
reasons discussed in Paper IV.  \iue\/ data points in the region $1215 \leq
\lambda \leq 1235$ \AA\/ are excluded due to contamination from
scattered solar Ly$\alpha$ photons.  The solid curves are the
parametrized fits to the data as determined by the SED-fitting
procedure discussed in \S \ref{secMETHOD}, and the shaded regions show
the 1-$\sigma$ uncertainty in the curves, based on Monte Carlo
simulations.  For comparison, the dash-dot curves show an estimate of
the average Galactic extinction curve from F99 (corresponding to $R(V)
= 3.1$).  Only the first panel of the figure is shown in the print edition
of the Journal.  The entire figure consists of 33 panels and is available
in the electronic edition.
\label{figATLAS}}
\end{figure}
\clearpage

\begin{figure}[ht]
\epsscale{0.8}
\plotone{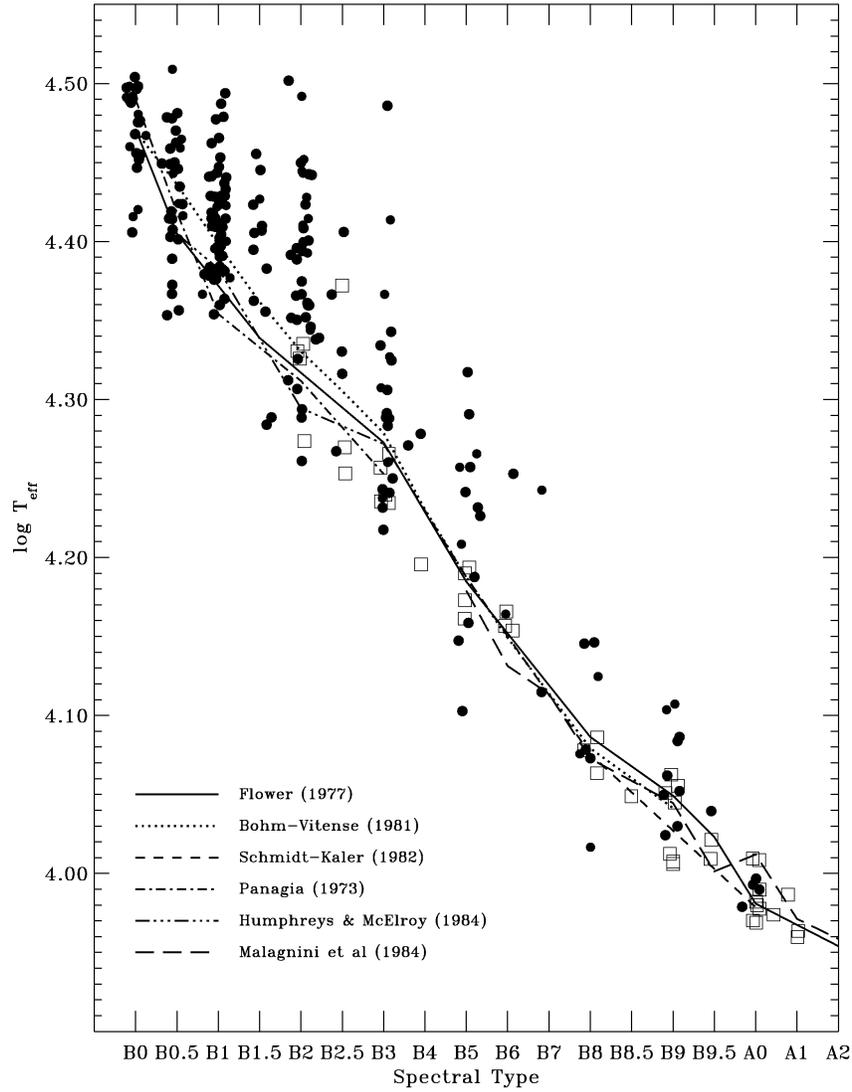}
\caption{$\log T_{eff}$ as a function of spectral type for the survey
stars of type B0 or later.  Luminosity class V or IV stars are shown by
filled circles; higher luminosity classes by open circles.  For
comparison, data for the 44 unreddened stars from the photometric
calibration study of Fitzpatrick \& Massa (2005b) are shown by the open
squares.  The various dotted, dashed, and solid lines show a number of
published spectral type vs. $T_{eff}$ calibrations, as indicated in the
figure. Small random horizontal offsets have been added to the data
points to increase their visibility.
\label{figSPTEFF}}
\end{figure}
\clearpage

\begin{figure}[ht]
\epsscale{0.8}
\plotone{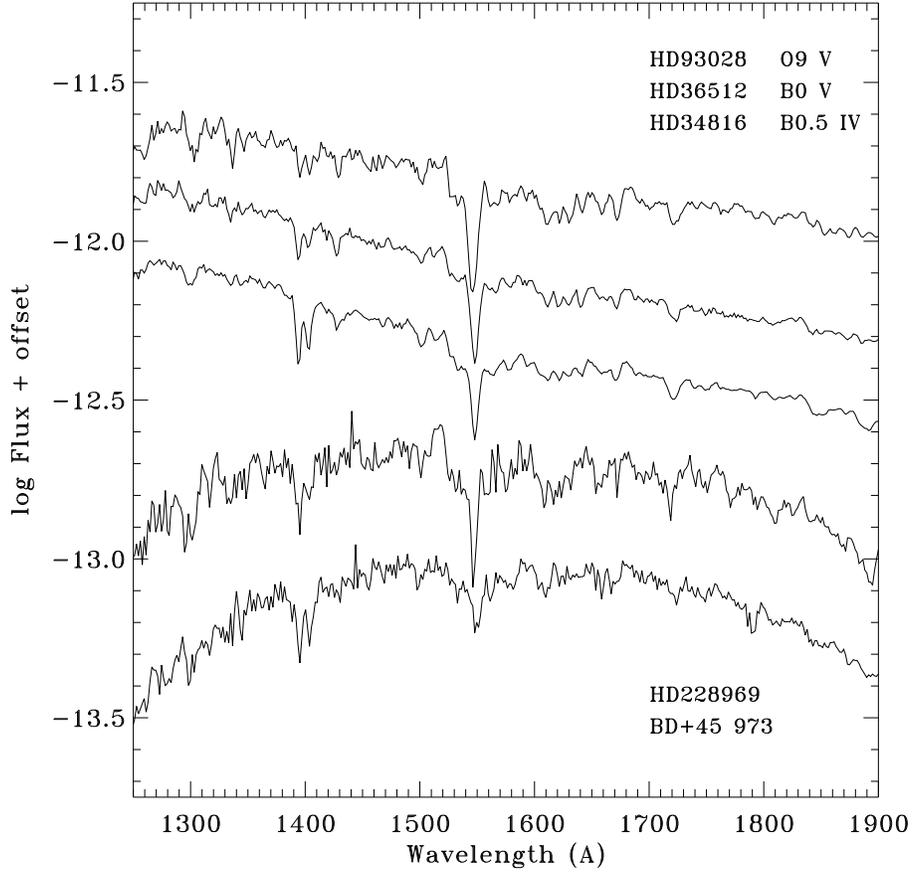}
\caption{UV spectra of two survey stars, HD228969 and BD+54 973,
compared with three early-type spectral classification standard stars.
The reddened survey stars are classified as types B2 II: and B3 V,
respectively, although the SED-fitting procedure indicates temperatures
in the neighborhood of 30000 K.  Comparison of the UV spectral features
with the classification standards indicates that these two stars are
likely misclassified members of earlier spectral classes.  Much of the
scatter in the \teff\/ vs. spectral type diagram in Figure
\ref{figSPTEFF} is probably the result of such classification
uncertainties.
\label{figCOMPARE}}
\end{figure}
\clearpage

\begin{figure}[ht]
\epsscale{0.8}
\plotone{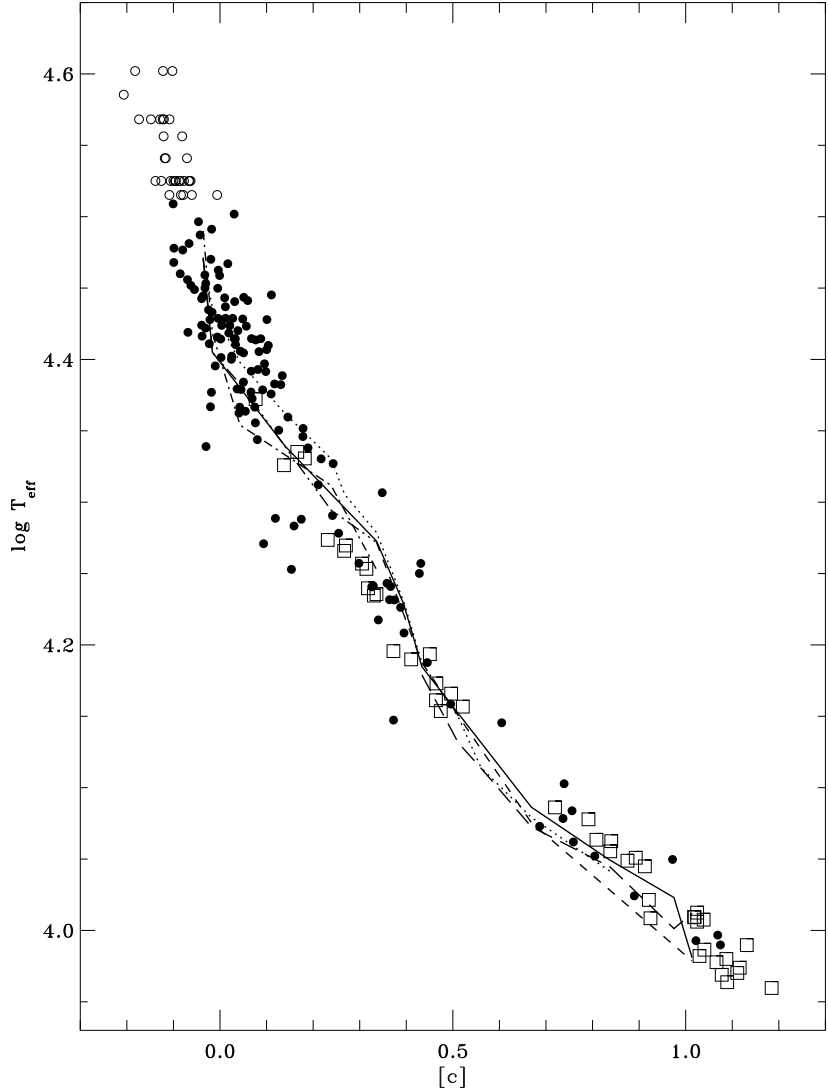}
\caption{\teff\/ as a function of the Str\"{o}mgren reddening-free
index $[c] \equiv c_1 - 0.20 (b-y)$, which measures the strength of the
Balmer jump, for 162 sample stars.  Symbols are the same as for Figure
\ref{figSPTEFF}, with the addition of the open circles which denote
O-type stars.  Also shown are the spectral type vs. \teff\/
calibrations from Figure \ref{figSPTEFF}, transformed into the \teff\/
vs. $[c]$ plane using Crawford's (1978) spectral type vs. $c_0$ and
$c_0$ vs. $(b-y)_0$ relations.  Note that Str\"{o}mgren photometry was not
used in modeling the SEDs of our survey stars.
 \label{figCBRACKET}}
\end{figure}
\clearpage

\begin{figure}[ht]
\epsscale{0.8}
\plotone{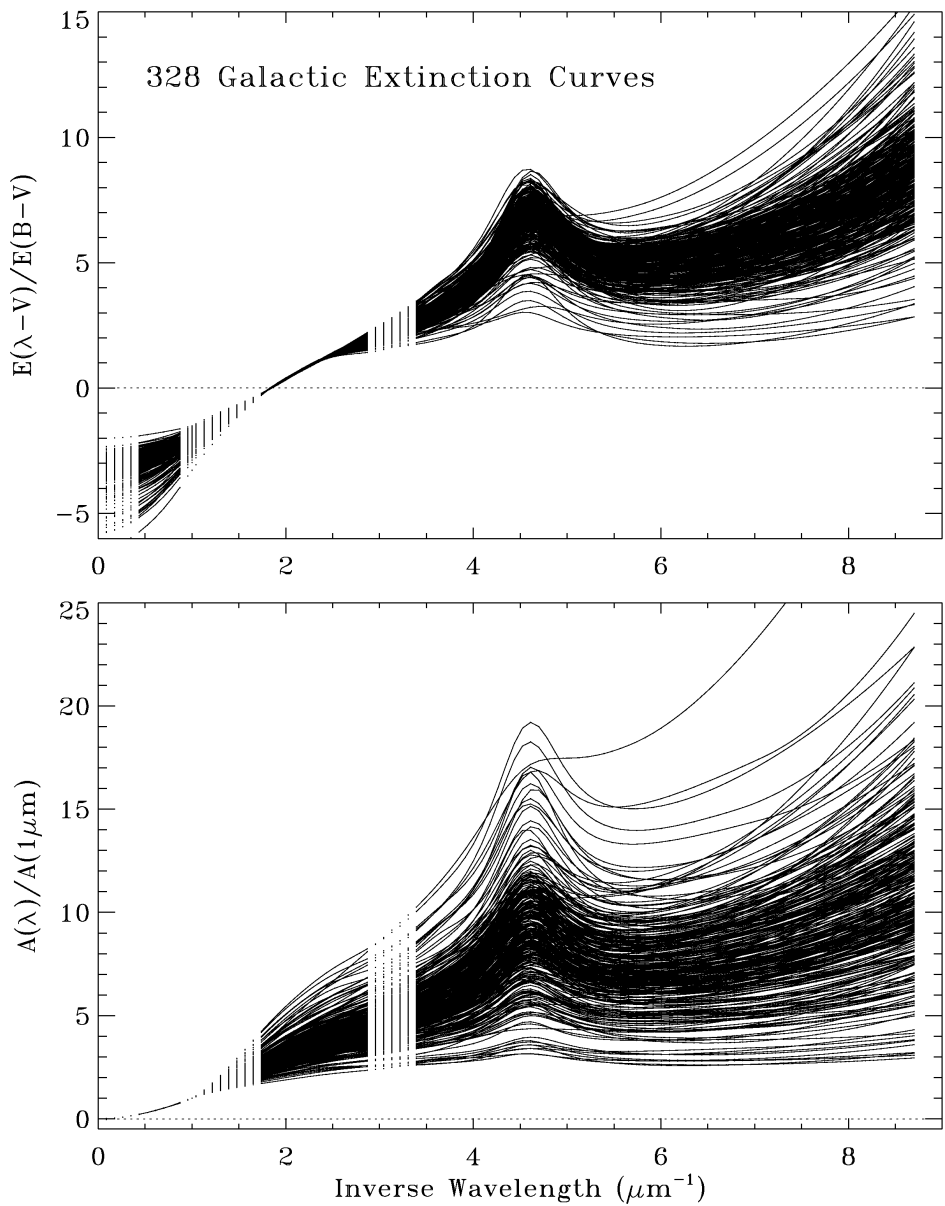}
\caption{Analytical representations of 328 Galactic UV-through-IR
extinction curves plotted in their native normalization
$E(\lambda-V)/E(B-V)$ ({top panel}) and transformed into an
IR-normalized form $A(\lambda)/A(1\/\mu{\rm m})$ ({bottom panel}).  The
curves, in their standard form, can be reproduced from the parameters
given in Table \ref{tabEXTINCT}.  The convergence of the curves in the
optical region (top panel) and the IR region (bottom panel) results
from their normalizations.  The value of $R(V) \equiv A(V)/E(B-V)$ for
each curve is the negative of its intercept at $\lambda^{-1} = 0 \;
\invmic$ in the $E(\lambda-V)/E(B-V)$ panel.
\label{figALLCURVES}}
\end{figure}
\clearpage

\begin{figure}[ht]
\epsscale{1.0}
\plotone{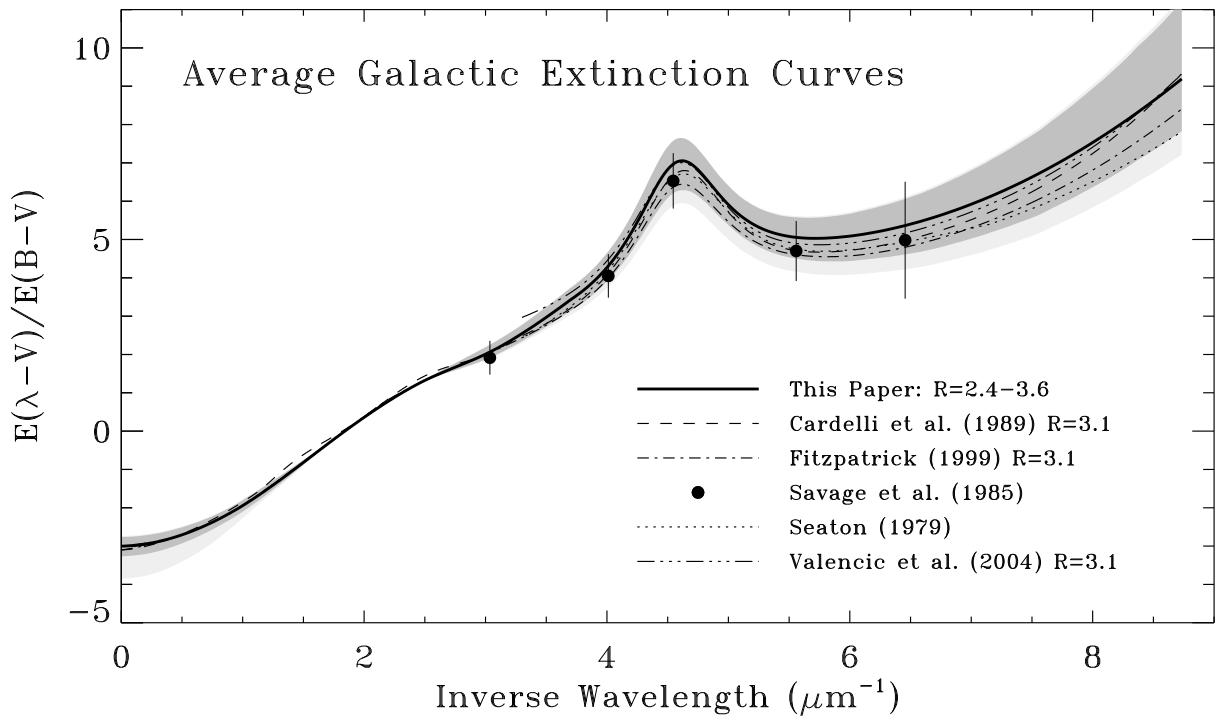}
\caption{An average UV-through-IR extinction curve for our sample
compared with other mean extinction curves.  The thick solid curve is
the mean for the 243 stars in our sample with $2.4 < R(V) < 3.6$, i.e.,
for those sightlines with $R(V)$ values indicative of the diffuse ISM.
The dark gray shaded region shows the sample variance about the mean
curve at all wavelengths.  The light grey shaded area shows the larger
variance that results when we include all 298 sightlines with measured
$R(V)$ values.  The mean curve can be reconstructed from the extinction
parameters listed in Table \ref{tabAVGCURVE}.  The dashed and dotted
curves show estimates of mean UV Galactic curves from the sources
indicated in the figure.  The large filled circles are means from the
\ans\/ satellite extinction catalog of Savage et al. (1985) for 800
stars with $\ebv \ge 0.20$ mag.  These measurements result from filter
photometry centered at wavelengths of 3000, 2500, 2200, 1800, and 1550
\AA.  Error bars on the \ans\/ data show the sample variances and
include the effects of spectral mismatch.
\label{figAVGCURVE}}
\end{figure}
\clearpage

\begin{figure}[ht]
\epsscale{1.0}
\plotone{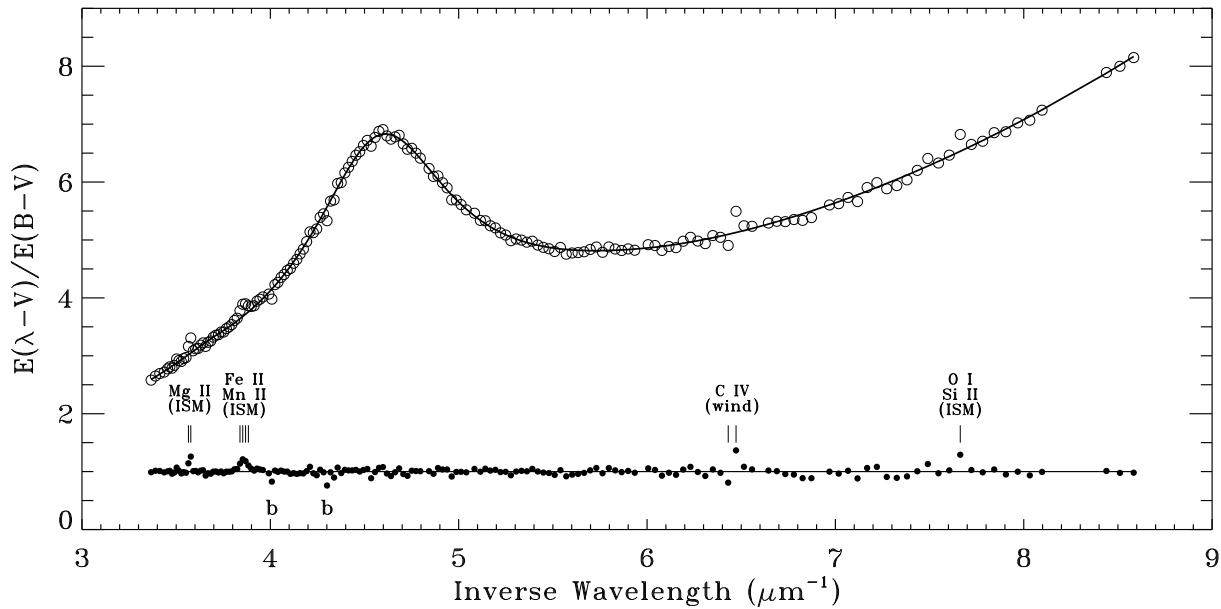}
\caption{Unweighted mean UV extinction curve for 318 survey stars (open
circles).  Ten stars (HD14092, BD+56 501, HD14321, HD18352, HD25443,
CD-42 4819, HD99872, HD326328, HD197702, and HD239710) were excluded
from the mean because they have incomplete \iue\/ spectra.  The
individual data points (shown in Figure \ref{figATLAS}) are typically
10 \AA\/ apart.  The solid curve is
a parametrized fit to the mean curve, using the formulation discussed
in \S \ref{secMETHOD}.  The $O-C$ residuals are shown as filled points;
they are offset from their zero mean for display.  A number of distinct
features are seen in the residuals and labeled in the figure.  These arise
from ISM gas-phase absorption lines, mismatch between the
C~{\sc iv}~$\lambda$1550 stellar wind lines in the O stars and the static
model SEDs, and known inadequacies in the \atlas\/ opacity distribution
functions (labeled ``b'' in the figure; see Paper IV).  No other credible
features are seen in the residuals.  The standard deviation of the
residuals about their mean value of zero (excluding the labeled points) is
$0.06\ebv$ mag, corresponding to $\sim0.02A(V)$ mag.  This figure
demonstrates both the intrinsic smoothness of UV extinction and the
ability of our parametrization scheme to reproduce the shape of UV
extinction curve to extremely high accuracy.
\label{figIUECURVE}}
\end{figure}
\clearpage

\topmargin -0.5in
\begin{figure}[ht]
\epsscale{0.9}
\plotone{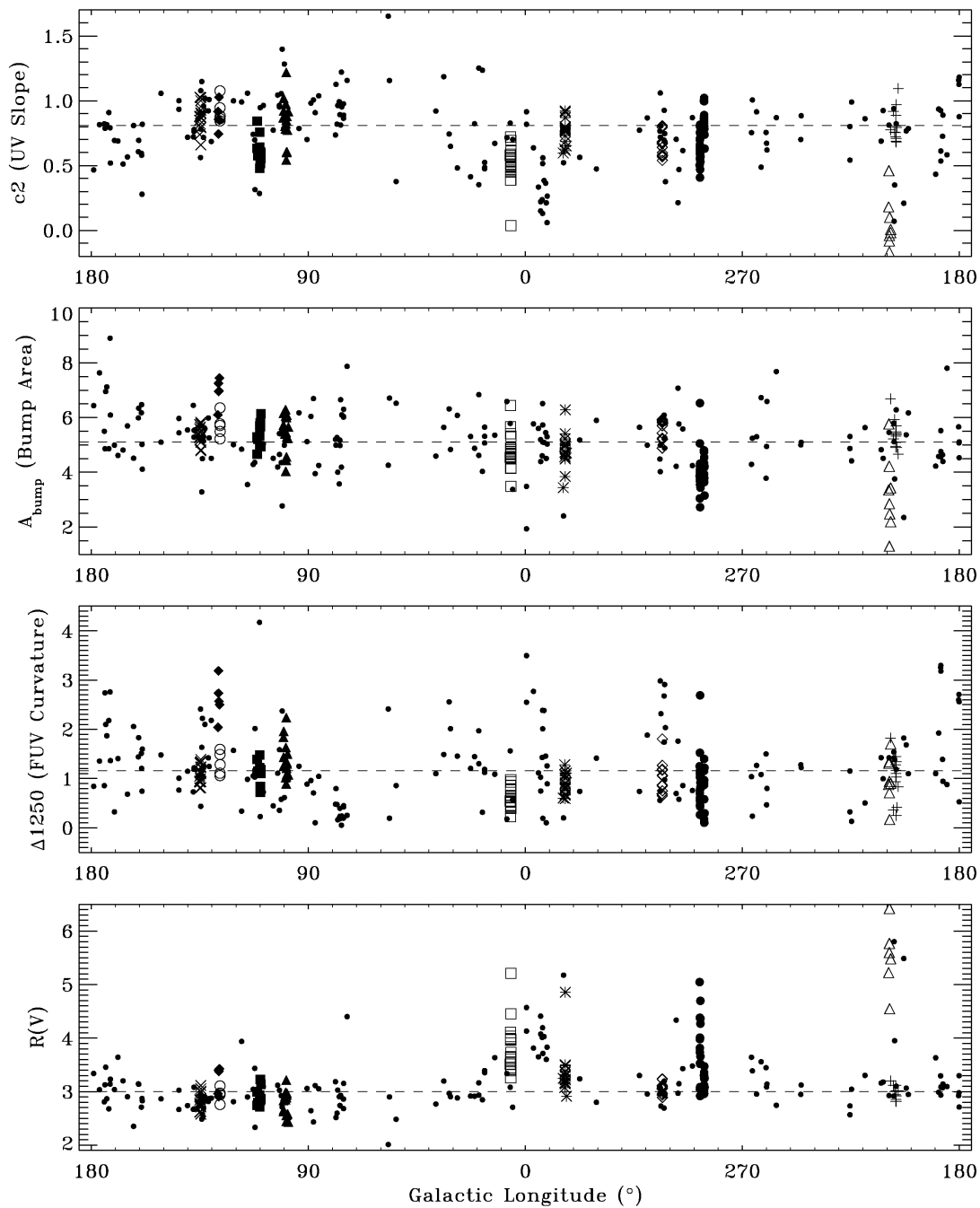}
\caption{Spatial trends in extinction properties.  The values of four
parameters which describe the shapes of interstellar extinction curves
are plotted against Galactic longitude for each of our survey
sightlines.  Small filled circles show field sightlines.  Other symbols
denote sightlines to clusters or associations for which five or more
members are included in our survey.  From left to right, the clusters
are: h \& $\chi$ Per (x's); NGC 457 (open circles); Cr 463 (filled
diamonds); Cep OB3 (filled squares); Trumpler 37 (filled triangles);
NGC 6530 (open squares); NGC 6231 (*'s); NGC 4755 (open diamonds);
Carina clusters (large filled circles, includes Trumpler 14 and 16, Cr
228, and NGC 3293); NGC 1977 (open triangles); and NGC2244 (+'s).  The
dashed lines show the parameter values corresponding to the diffuse
mean ISM curve from Figure \ref{figAVGCURVE}.  
\label{figLONGITUDE}}
\end{figure}
\clearpage

\begin{figure}[ht]
\epsscale{0.9}
\plotone{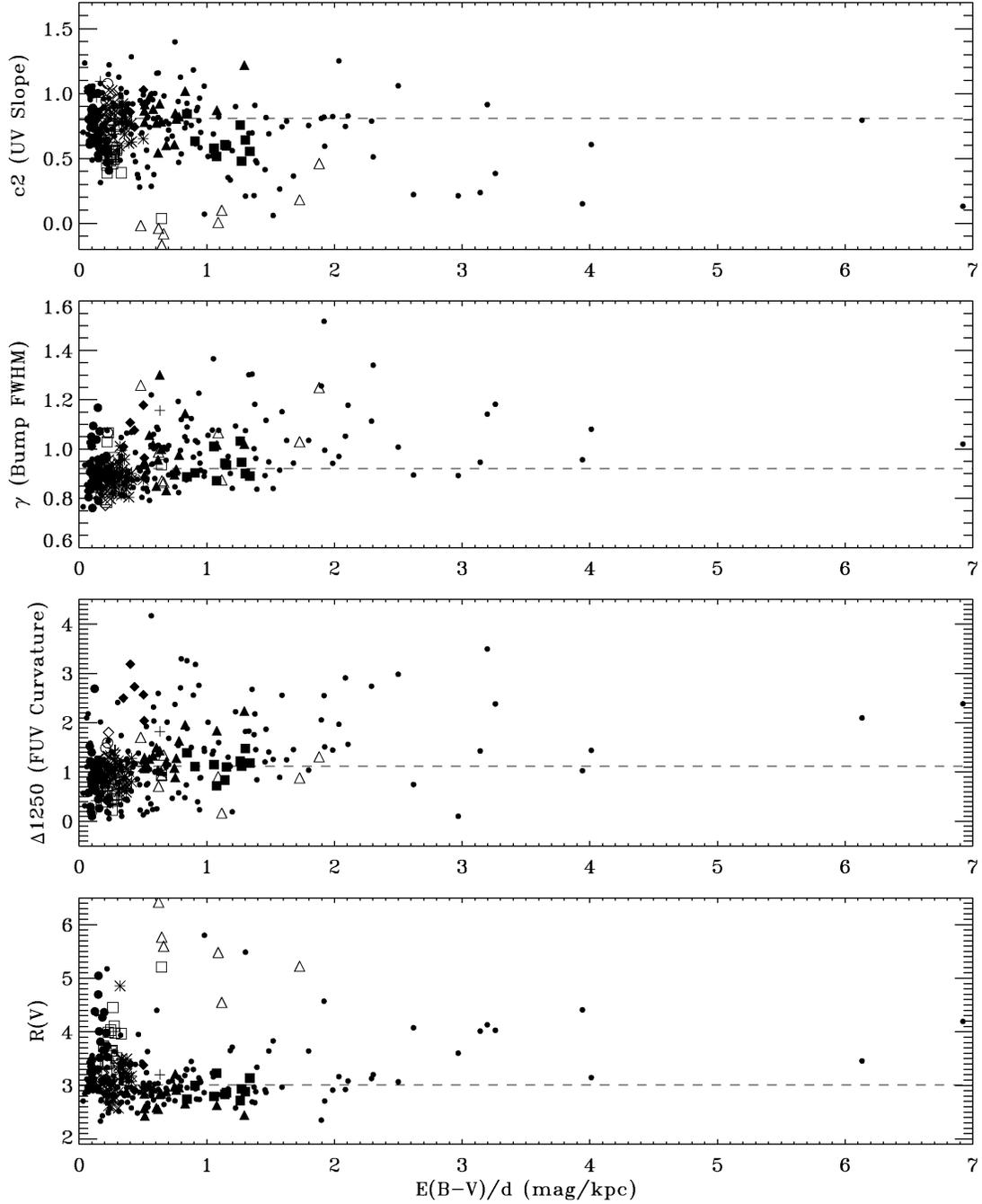}
\caption{Density trends in extinction properties.  The values of four
parameters which describe the shapes of interstellar extinction curves
are plotted against \ebv$/d$ for our survey sightlines.
Small filled circles are field sightlines.  Other symbols (same as in
Figure~\ref{figLONGITUDE}) denote sightlines to clusters or associations
containing five or more survey stars.   The dashed lines show the parameter
values corresponding to the diffuse mean ISM curve in
Figure~\ref{figAVGCURVE}.
\ref{figAVGCURVE}.
\label{figDENSITY}}
\end{figure}
\clearpage

\topmargin 0.0in
\begin{figure}[ht]
\epsscale{1.1}
\plotone{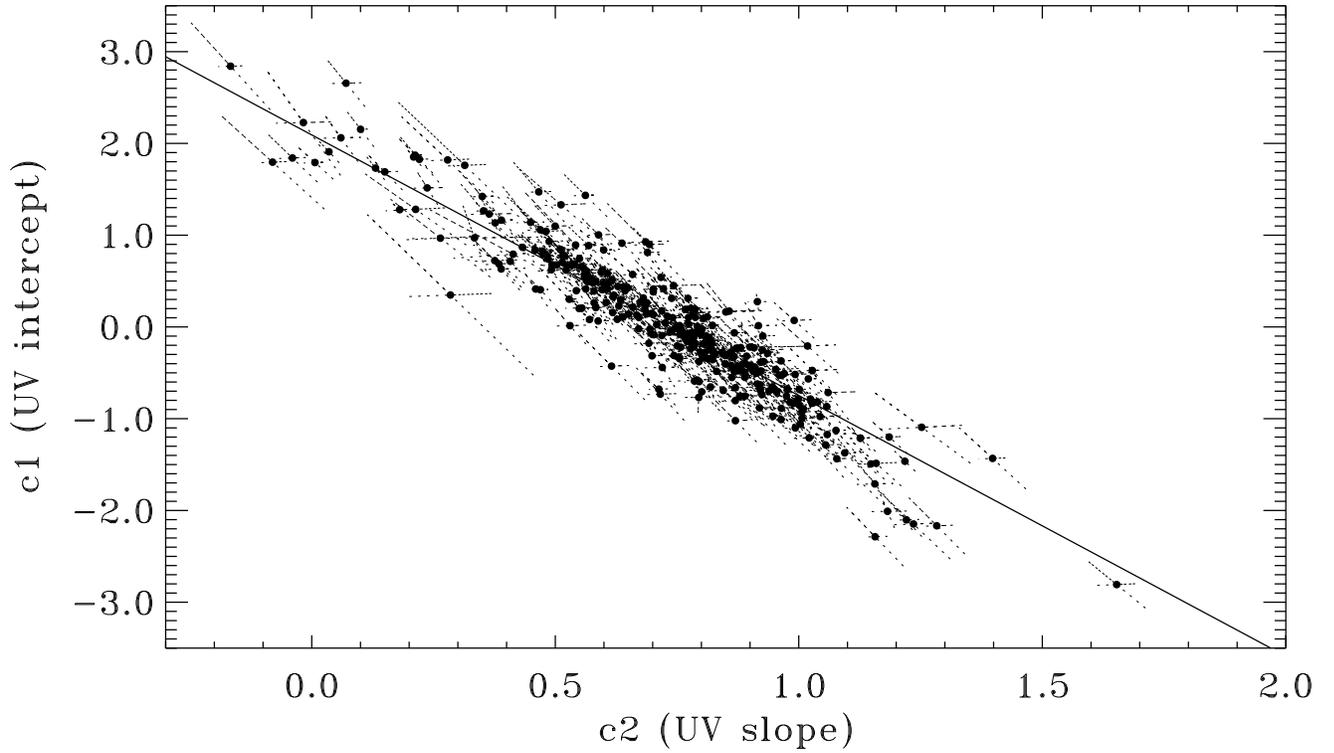}
\caption{Plot of the slope $c_2$ vs. the intercept $c_1$ for the
linear component of UV extinction.  The uncertainties in $c_1$ and
$c_2$ are strongly correlated, as indicated by the 1-$\sigma$ error
bars (dotted lines).  The orientation of the error bars was determined
by fitting ellipses to the loci of results from the Monte Carlo error
simulations. The individual sets of error bars are not orthogonal to
each other because the scales of the x- and y-axes are different.  The
solid line is a weighted linear fit to the data, designed to minimize
the residuals in the direction perpendicular to the fit. It is given by
$c_1 = 2.09 - 2.84 c_2$.
\label{figC1C2}}
\end{figure}
\clearpage

\begin{figure}[ht]
\epsscale{1.1}
\plotone{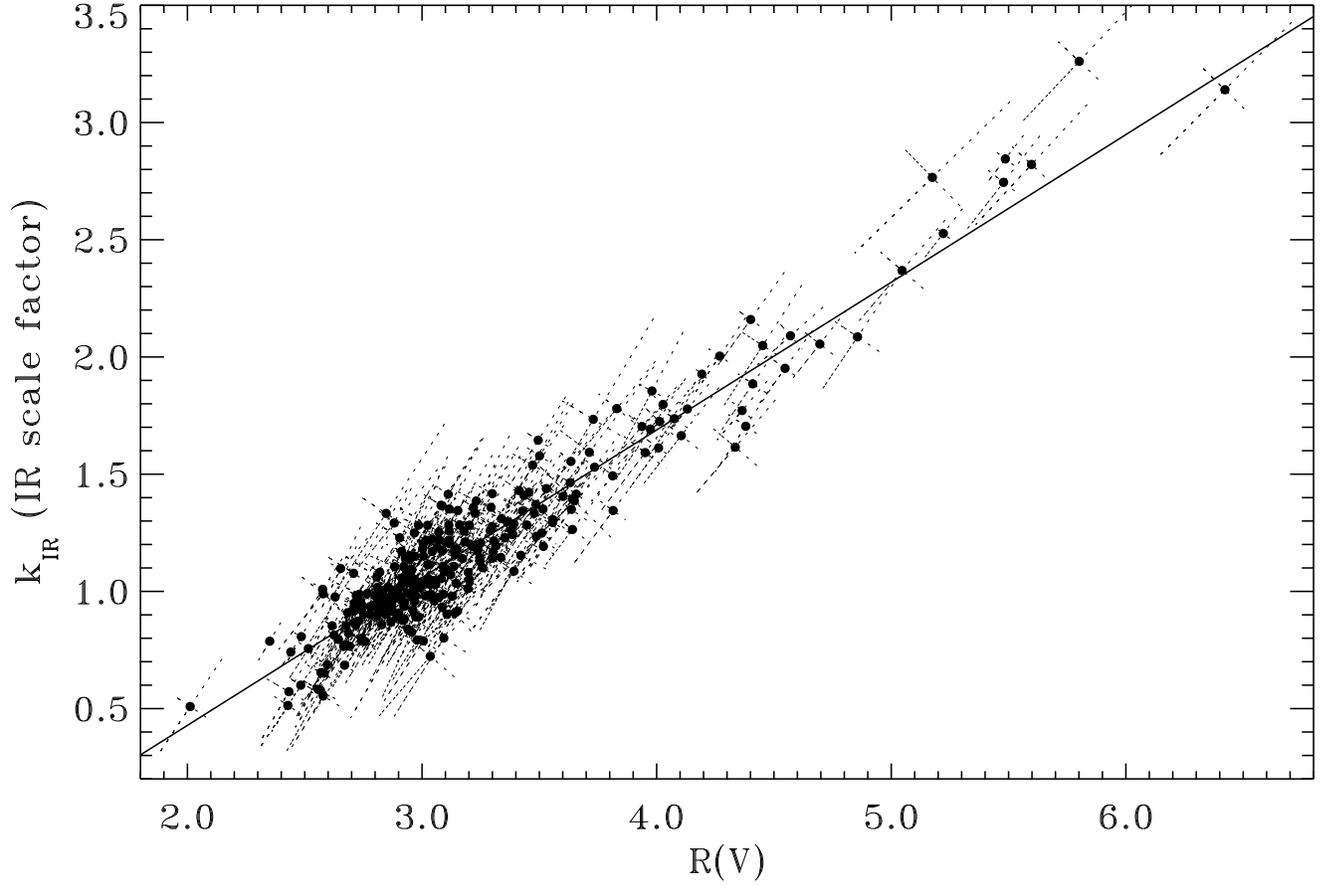}
\caption{Plot of $R(V)$ vs. the IR scale factor $k_{IR}$ (see
Eq.~[\ref{eqnIR}]).  The uncertainties in $R(V)$ and $k_{IR}$ are
strongly correlated, as indicated by the 1-$\sigma$ error bars (dotted
lines).  The orientation of the error bars was determined by fitting
ellipses to the loci of results from the Monte Carlo error
simulations.  The solid line is a weighted linear fit to the data,
designed to minimize the residuals in the direction perpendicular to
the fit.  It is given by $k_{IR} = -0.83 + 0.63 R(V)$.
\label{figRKIR}}
\end{figure}
\clearpage

\begin{figure}[ht]
\epsscale{1.0}
\plotone{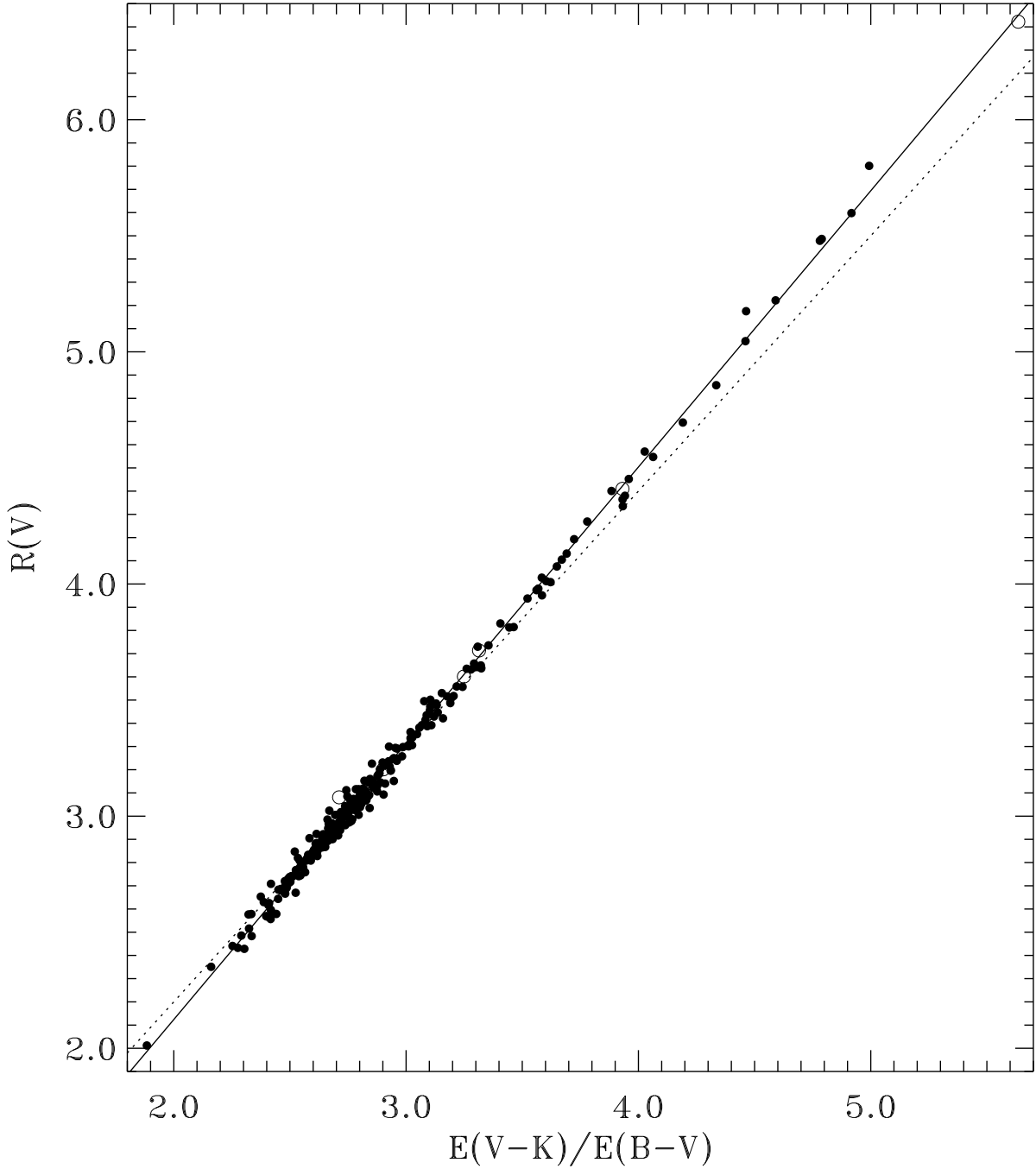}
\caption{Plot of $E(V-K)/E(B-V)$ vs. $R(V)$ for sightlines with \tmass\/
$K$-band photometry (filled circles) and Johnson $K$-band photometry
(open circles).  The dotted line shows the widely-used relation $R(V) =
1.1E(V-K)/E(B-V)$.  The solid line is a linear fit to the \tmass\/ data,
minimizing the residuals in the direction perpendicular to the fit.  It
is given by $R(V) = -0.26 + 1.19E(V-K)/E(B-V)$.
\label{figEVMINK}}
\end{figure}
\clearpage

\begin{figure}[ht]
\epsscale{1.1}
\plotone{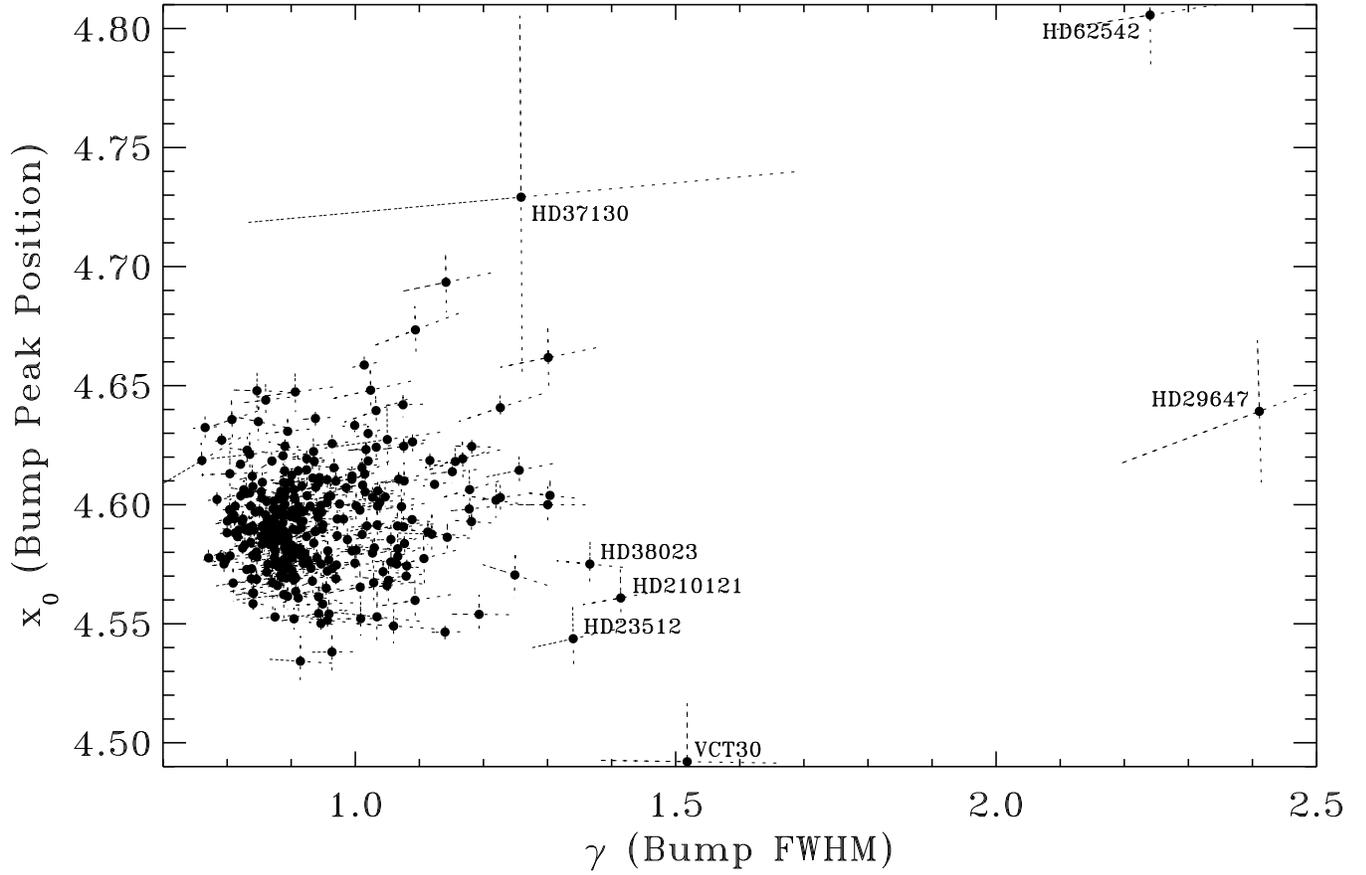}
\caption{Plot of the FWHM of the 2175 \AA\/ bump $\gamma$ vs. its
central position, $x_0$, both in units of \invmic.  Dotted lines
show the 1-$\sigma$ error bars, revealing the uncertainties in
$\gamma$ and $x_0$ to be generally uncorrelated.  The orientation of
the error bars was determined by fitting ellipses to the loci of
results from the Monte Carlo error simulations.  The individual sets of
error bars are not orthogonal to each other because the scales of the
x- and y-axes are different.  A number of sightlines which deviate most
from the main distribution have been labeled.
\label{figGAMMAX0}}
\end{figure}
\clearpage

\begin{figure}[ht]
\epsscale{1.1}
\plotone{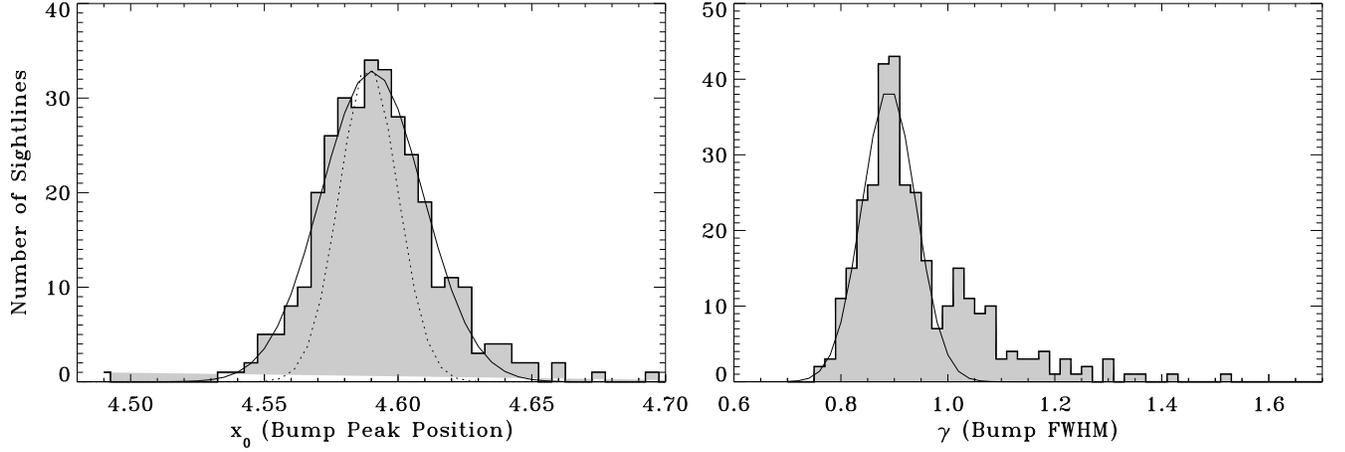}
\caption{2175 \AA\/ bump statistics.  The left panel shows a histogram
of the  distribution of the 2175 \AA\/ bump peak positions for our
sample.  The smooth curve is a Gaussian fit centered at $x_0 = 4.5903
\; \invmic$ (2178.5 \AA) with a width of $\sigma = 0.0191 \; \invmic$
(9.1 \AA).  The RMS value of the measurement errors in $x_0$ is $\pm
0.0058  \; \invmic$ ($\pm 2.8 \; {\rm \AA}$).  The dashed curve shows a
Gaussian fit to the distribution of bump positions for 154 stars
located in the 13 open clusters with more than 5 stars in the survey.
For each cluster, a mean value of $x_0$ was computed and the
distributions constructed relative to the mean (see \S
\ref{secEXT_X0GAMMA}).  The width of the cluster Gaussian is given by
$\sigma = 0.011 \; \invmic$ (5.2 \AA).  The right panel shows the
distribution of bump FWHM values.  The smooth curve is a Gaussian fit
to the main peak in the distribution.  Its central position corresponds
to $\gamma = 0.890 \; \invmic$ and its width is $\sigma = 0.050 \;
\invmic$.  The RMS measurement error in $\gamma$ is $\pm 0.031 \;
\invmic$.
\label{figBUMPSTATS}}
\end{figure}
\clearpage

\begin{figure}[ht]
\epsscale{1.0}
\plotone{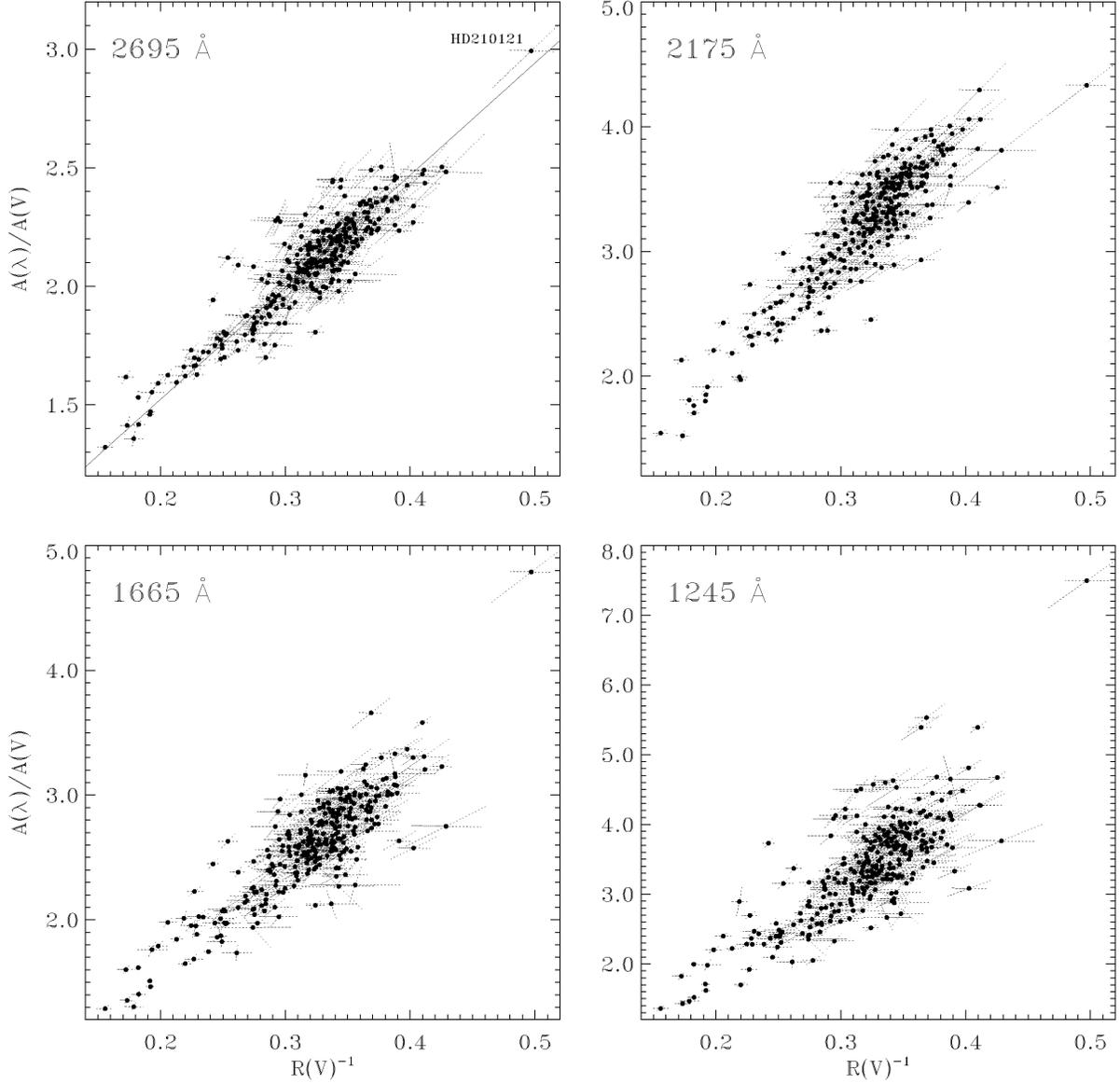}
\caption{Plots of the normalized extinction values $A(\lambda)/A(V)$
vs. $R(V)^{-1}$ at four different UV wavelengths. $A(\lambda)/A(V)$ is
derived from the measured values of $E(\lambda-V)/E(B-V)$ by using
Equation (\ref{eqnCCM2}).  1-$\sigma$ error bars are
shown, based on our Monte Carlo simulations.  The error bars, which are
not orthogonal to each other because the scales of the x- and y-axes
are different, show the clear correlation in the uncertainties for
$A(\lambda)/A(V)$ and $R(V)^{-1}$.  The straight line in the 2695 \AA\/
panel is a weighted linear fit to the data, which minimizes the
scatter perpendicular to the relation.  It is given by  $A(2695 \;
{\rm \AA})/A(V) = 0.58 + 4.73 \; R(V)^{-1}$. 
\label{figRVAV}}
\end{figure}
\clearpage

\begin{figure}[ht]
\epsscale{1.0}
\plotone{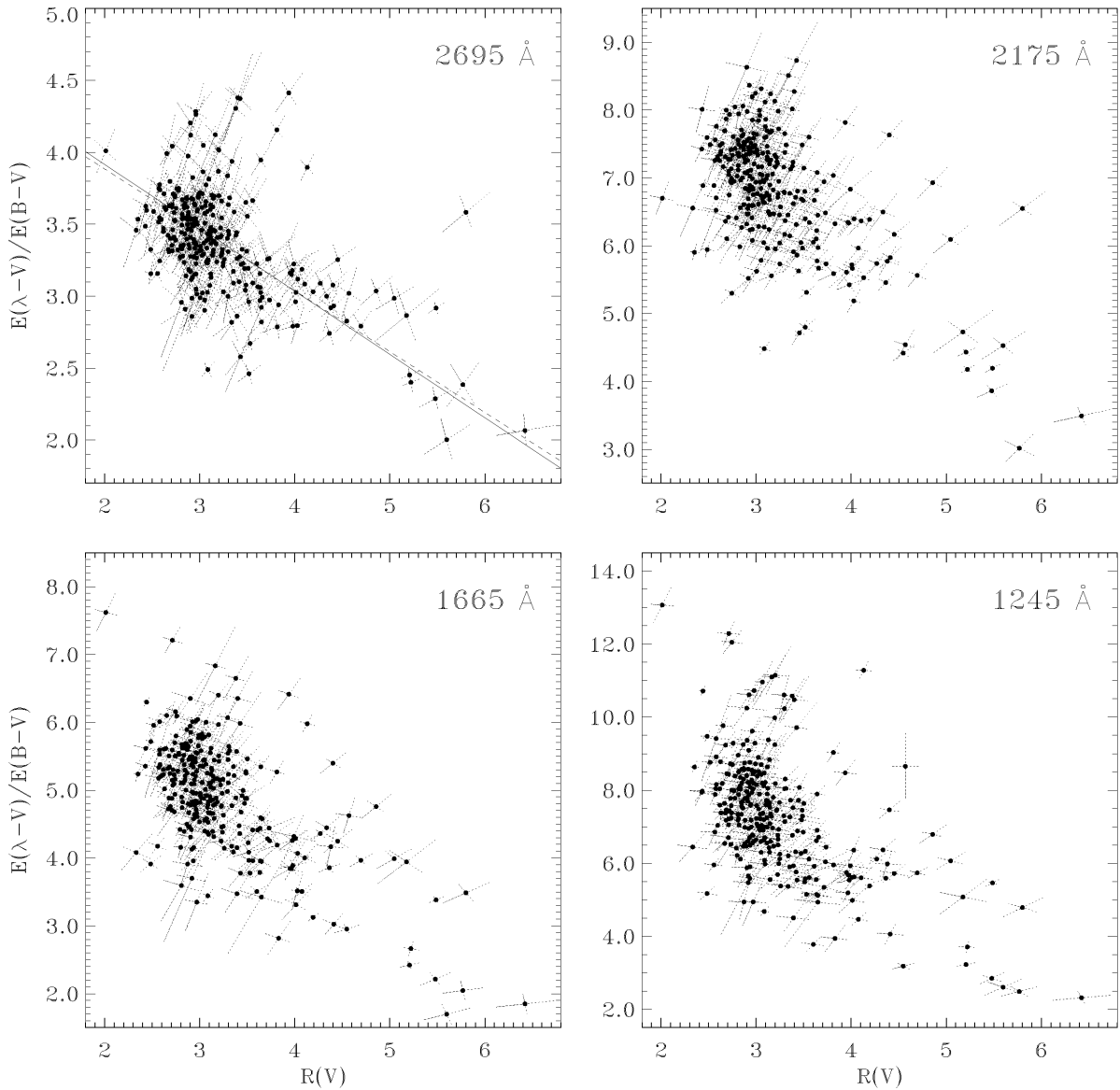}
\caption{Plots of the measured extinction values $E(\lambda-V)/E(B-V)$
vs.\ $R(V)$ at four different UV wavelengths.  1-$\sigma$ error bars
are shown, based on our Monte Carlo simulations as in previous
figures.  Any true relationships seen in plots of $A(\lambda)/A(V)$
vs.\ $R(V)^{-1}$ are preserved in plots of $E(\lambda-V)/E(B-V)$
vs.\ $R(V)$, but with the great benefit that the correlations between
the errors are greatly reduced.  The solid line in the 2695 \AA\/ panel
is a weighted linear fit to the data, which minimizes the scatter
perpendicular to the relation.  This fit is $E(2695 \; {\rm \AA} \;
-V)/E(B-V) = 4.80 - 0.44 \; R(V)$.  The nearly coincident dashed line
is the fit in Figure~\ref{figRVAV} transformed by Equation
(\ref{eqnCCM3}).  This fit is $E(2695 \; {\rm \AA} \; -V)/E(B-V) = 4.73
- 0.42 \; R(V)$.
\label{figRVKLAM}}
\end{figure}
\clearpage

\topmargin -0.5in
\begin{figure}[ht]
\epsscale{0.8}
\plotone{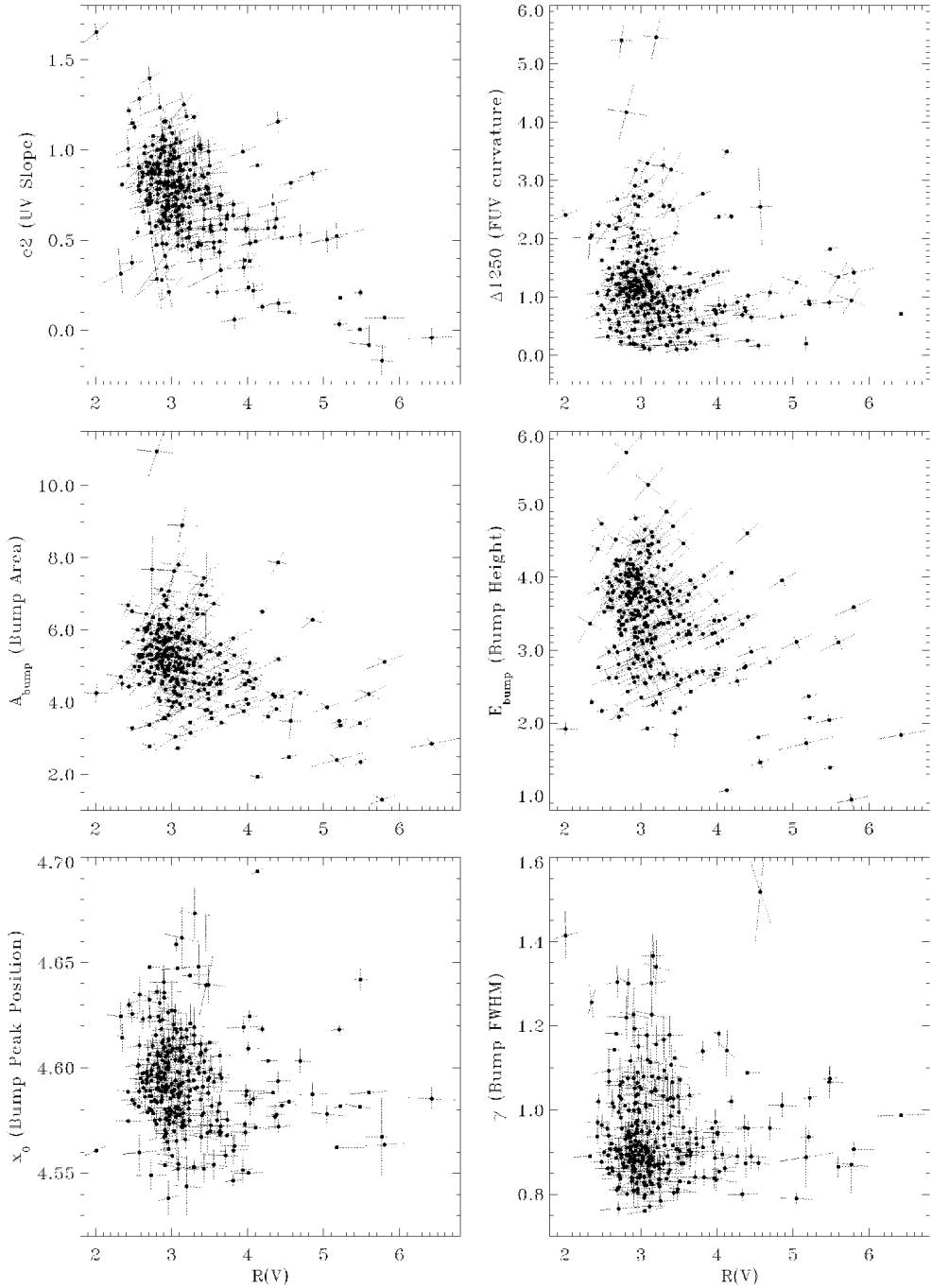}
\caption{Plots of UV fit parameters (and derived quantities) vs.
$R(V)$.  1-$\sigma$ error bars are shown, based on our Monte Carlo
simulations as in previous figures.  The quantity $\Delta$1250 is the
difference between the observed value of $E(\lambda-V)/E(B-V)$ at 1250
\AA\/ (8.0 \invmic) and an extrapolation of the linear plus bump
components of UV extinction.  It thus measures the strength of the FUV
curvature and is computed by $\Delta1250 = c_4(8.0-c_5)^2$ (see Eq.
[\ref{eqnFMFUNC}]).  $A_{bump} \equiv \pi \; c_3/(2\gamma)$ is the area
under the 2175 \AA\/ bump.  $E_{bump} \equiv c_3/\gamma^2$ is the
height of the 2175 \AA\/ bump above the linear background extinction.
\label{figRVFM}}
\end{figure}
\clearpage

\topmargin -0.0in
\begin{figure}[ht]
\epsscale{0.8}
\plotone{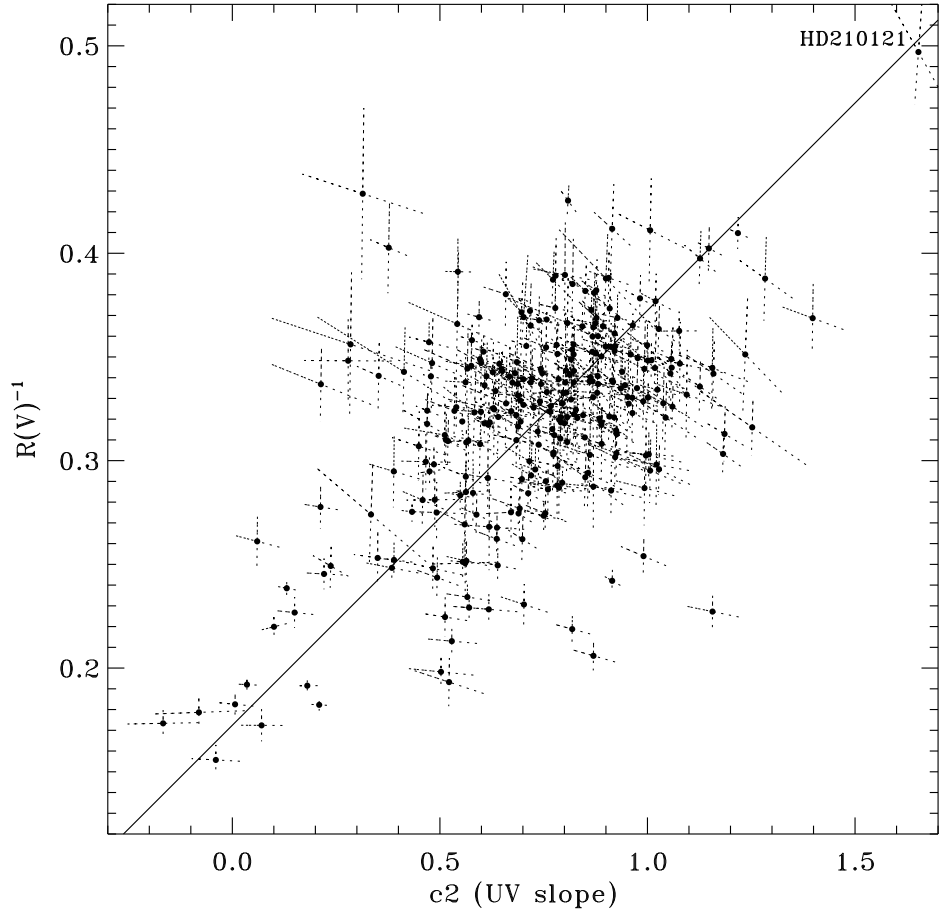}
\caption{Plot of the slope of the UV linear component $c_2$ vs.
$R(V)^{-1}$.  1-$\sigma$ error bars are shown, based on our Monte Carlo
simulations as in previous figures.  The solid line is a weighted
linear fit which minimizes the residuals in the direction perpendicular
to the fit.  It is given by $R(V)^{-1} = 0.17 + 0.20 \; c_2$.
\label{figRVC2}}
\end{figure}

\end{document}